\documentclass[aps,prb,onecolumn,superscriptaddress,nofootinbib,longbibliography]{revtex4-1} 

\usepackage[pdftex,colorlinks,hyperindex,plainpages=false,bookmarksopen,bookmarksnumbered,pdfusetitle,allcolors={blue}]{hyperref}

\usepackage{longtable}
\usepackage{morefloats}
\usepackage{color}
\usepackage{bm}
\usepackage{graphicx,amsfonts,amsbsy}
\usepackage{amsmath,amsfonts,amsthm,amssymb}
\usepackage{appendix}
\usepackage{makeidx}
\usepackage{url}
\usepackage{verbatim}

\usepackage[rightcaption]{sidecap}
\usepackage{array}
\usepackage{booktabs}
\usepackage{multirow}
\usepackage{tabularx}
\usepackage{cancel,ulem}
\usepackage{mathrsfs}
\usepackage{soul,xcolor}
\usepackage{times}
\usepackage{txfonts}

\newcommand{\bs}[1]{\boldsymbol{#1}}

\begin{document}
\setstcolor{red}

\title{Thermal squeezing and nonlinear spectral shift of magnons in antiferromagnetic insulators}

\author{Mahroo Shiranazei}
\affiliation{Division of Materials Theory, Department of Physics and Astronomy, Uppsala University, Box 516, SE-75120 Uppsala, Sweden}
\author{Roberto E. Troncoso}
\affiliation{Center for Quantum Spintronics, Department of Physics, Norwegian University of Science and Technology, NO-7491 Trondheim, Norway}
\affiliation{School of Engineering and Sciences, Universidad Adolfo Ib\'a\~nez, Santiago, Chile}

\author{Jonas Fransson}
\affiliation{Division of Materials Theory, Department of Physics and Astronomy, Uppsala University, Box 516, SE-75120 Uppsala, Sweden}
\author{Arne Brataas}
\affiliation{Center for Quantum Spintronics, Department of Physics, Norwegian University of Science and Technology, NO-7491 Trondheim, Norway}
\author{Alireza Qaiumzadeh}
\affiliation{Center for Quantum Spintronics, Department of Physics, Norwegian University of Science and Technology, NO-7491 Trondheim, Norway}

\begin{abstract}
We investigate the effect of magnon-magnon interactions on the dispersion and polarization of magnon modes in collinear antiferromagnetic (AF) insulators at finite temperatures. In two-sublattice AF systems with uniaxial easy-axis and biaxial easy-plane magneto-crystalline anisotropies, we implement a self-consistent Hartree-Fock mean-field approximation to explore the nonlinear thermal interactions. The resulting nonlinear magnon interactions separate into two-magnon intra- and interband scattering processes. Furthermore, we compute the temperature dependence of the magnon bandgap and AF resonance modes due to nonlinear magnon interactions for square and hexagonal lattices. In addition, we study the effect of magnon interactions on the polarization of magnon modes. We find that although the noninteracting eigenmodes in the uniaxial easy-axis case are circularly polarized, but in the presence of nonlinear thermal interactions the U(1) symmetry of the magnon Hamiltonian is broken. The attractive nonlinear interactions squeeze the low energy magnon modes and make them elliptical. In the biaxial easy-plane case, on the other hand, the bare eigenmodes of low energy magnons are elliptically polarized but thermal nonlinear interactions squeeze them further. 
Direct measurements of the predicted temperature-dependent AF resonance modes and their polarization can be used as a tool to probe the nonlinear interactions. Our findings establish a framework for exploring the effect of thermal magnon interactions in technologically important magnetic systems, such as magnetic stability of recently discovered two-dimensional magnetic materials, coherent transport of magnons, Bose-Einstein condensation of magnons, and magnonic topological insulators.
\end{abstract}

\maketitle

\section{Introduction}\label{Intro}
Antiferromagnetic (AF) systems occupy a central position at the frontiers of modern spintronics due to their functional potential in future energy-efficient and ultrafast information and communication nanotechnologies \cite{MartiNM2014,KriegnerNC2016,Lebrun_2018,NvemecNP2018,GepragsJAP2020}. AF ordered systems are abundant materials compared to ferromagnets and range from collinear to noncollinear and uniaxial to multi-axis materials.
Current research centers on exploiting their most remarkable properties as the lack of parasitic stray fields, THz spin dynamics, and polarization degree of freedom of magnons \cite{BaltzRMP2018,PhysRevB.97.020402,ChumakNC2014,
BowlanJPD2018,JungwirthNN2016, PhysRevLett.119.177202}. Of particular interest are the electrical insulator class of AF materials \cite{ChumakNP2015,GomonayLTP2014,DanielsNJP2015,KhymynPRB2016,JungwirthNN2016,Gomonay2017,NishitaniAPL2010,NishitaniPRB2012,Lebrun_2020}, where Joule-heating losses caused by itinerant electrons are absent. Magnons, the low-energy bosonic excitations of magnetically ordered materials, can carry spin angular momentum in AF insulators across large distances, as recently shown\cite{Lebrun_2020,Lebrun_2018,RezendeJAP2019}. 

Magnonics is an emerging field in spintronics and magnetic-based nanotechnology \cite{ChumakNP2015} that exhibits various features from quantum coherent phenomena \cite{BunkovJP2010,DzyapkoPRB2017,TupitsynPRL2008, RezendePRB2009}, chiral magnonic spin transport, magnonic crystals \cite{TroncosoPRB2015,ZakeriJP2020,ChumakJPD2017,Singh2021}, and ultrafast magnetization dynamics \cite{KirilyukRMP2010}.
Nonlinear magnon interactions are ubiquitous and inevitable in magnetically ordered materials. Although in classical magnets nonlinear magnon-magnon interactions are weak at low temperatures, their effects significantly influence various phenomena. These phenomena are relevant in spin transport \cite{WangPRB2020,BayrakciPRL2013} through magnon relaxation and spin conductivity, as well as close to magnetic phase transitions. Moreover, nonlinear magnon interactions are essential to stabilize magnetic droplets \cite{MohseniS2013} and Bose-Einstein condensation of magnons \cite{MohseniNJP2020,TupitsynPRL2008, ArakawaPRB2019,GiamarchiNP2008}. It is also an important ingredient in emergent magnonic topological phases \cite{MookArXiv2020}. A recent study demonstrated that nonlinear magnon interactions of Dirac magnons in honeycomb ferromagnets strongly renormalize the magnon band structure and magnon lifetimes \cite{PhysRevX.8.011010}.  
In AF systems, nonlinear magnon interactions also have significant effects even at low temperatures due to zero-point fluctuations \cite{ZhitomirskyRMP2013}. Additionally, two-magnon spectra of AF systems are strongly influenced by magnon-magnon interactions  \cite{ElliottJPC1969}.

The recent discovery of two-dimensional (2D) materials establishes a new platform for the study of nonlinear magnon interactions in different lattice structures \cite{BayrakciPRL2013, PhysRevB.101.064416, niyazi2021antiferromagnetic,  PhysRevB.104.064435, LI201828}.
In low-dimensional magnetic systems, the presence of a finite bandgap in the low-energy magnon dispersion is an essential ingredient for the stability of long-range magnetic order at finite temperature and overcoming the bottleneck imposed by the Mermin-Wagner-Hohenberg theorem \cite{Mermin, Hohenberg}. Therefore, the effect of nonlinear magnon interactions on the magnetic resonance frequency and magnon dispersion is a key issue in the study of long-range order stability of 2D magnetic systems. 

There are already several primary works which investigate nonlinear magnon interactions in AF systems within a mean-field approach. In Refs. \onlinecite{castilla1991spin, igarashi19921}, the effect of nonlinear interactions on magnon dispersion of a square lattice Heisenberg antiferromagnet were studied via a Green's function method at zero temperature. In Ref. \onlinecite{an2001softened} effect of nonlinear thermal magnon interactions on spin-wave excitation and sublattice magnetization in a 3D anisotropic quantum Heisenberg antiferromagnet with a tuneable interplanar coupling has been studied via an equation of motion formalism within a mean-field approach. For a recent review on AF systems look at Ref. \onlinecite{RezendeJAP2019}.

Nonlinear magnon interactions, even in simple AF systems, are very complicated. To the best of our knowledge, a systematic investigation of nonlinear interactions on spectrum and polarization of thermal magnons in AF materials remains absent \cite{Liu_1992}. In AF systems, the underlying crystal lattice structure affects the magnetic ground states and interactions, and therefore, one needs to study each case separately \cite{PhysRevB.79.144416}. 

In this paper, we first develop a formalism to compute the nonlinear interactions of thermal magnons in both uniaxial easy-axis and biaxial easy-plane AF insulators in arbitrary dimensions using a self-consistent Hartree-Fock mean-field theory.
Next, we investigate the effect of thermal nonlinear interactions on bare magnon dispersion and AF resonance frequencies, which are directly accessible experimentally \cite{PhysRevLett.119.047201}. Later, we investigate the effect of nonlinear interactions on the polarization of magnon modes.
For numerical calculations, We consider collinear AF materials with easy-axis and easy-plane magnetic anisotropies in square and hexagonal lattices. We study these systems as prototypes of recently discovered 2D AF materials to explore the stability of these systems against thermal fluctuations.

We have organized this paper as follows. In Sec.\ \ref{model}, we introduce the spin Hamiltonian and review the linear spin-wave theory. In Sec.\ \ref{sec:interaction}, we present a Hartree-Fock mean-field theory for thermal magnon interactions and compute mean-field coefficients. In Sec.\ \ref{renormalization}, we compute the nonlinear spectral shift of magnons in easy-axis and easy-plane AF systems and present numerical calculations for two specific lattice models. In Sec.\ \ref{polarization}, we investigate the effect of nonlinear thermal interactions on the polarization of magnons. We conclude in Sec. \ref{sec:conclusion} with a discussion of our results. In the appendices, we present the technical details of the calculations.

\section{Spin Hamiltonian and its bosonic representation}\label{model}
We describe a generic two-sublattice AF insulator with the local spin field $\mathbf{S}_{\mathcal{A}(\mathcal{B})}$ in sublattice $\mathcal{A}$ ($\mathcal{B}$). The effective AF spin Hamiltonian reads,  \cite{BaltzRMP2018,RezendeJAP2019},
\begin{align}\label{Hspin}
H_{s} = J \sum_{\langle i j \rangle} \mathbf{S}_i \cdot \mathbf{S}_j + \sum_{i} \mathcal{K}_x ( S^x_i)^2 - \sum_{i} \mathcal{K}_z( S^z_i)^2,
\end{align}
where $\langle ij \rangle$ represents summation over nearest-neighbor sites, $i$ and $j$, and we consider the local spins have the same length $S$ at each sublattice $|\mathbf{S}_{i \in \{\mathcal{A}\}}|=|\mathbf{S}_{i \in \{\mathcal{B}\}}|=S$.
The Hamiltonian includes the isotropic quantum AF Heisenberg exchange interaction, $J>0$, and single-ion magnetic anisotropies, characterized by  $\mathcal{K}_x > 0$ and $\mathcal{K}_z > 0$, which represent the hard and easy magnetic axes, respectively. In the uniaxial easy axis case, i.e., $\mathcal{K}_x = 0$, both the spin Hamiltonian and the ground state have $U(1)$ symmetry while in the biaxial easy-plane case this symmetry is broken. The absence of $U(1)$ symmetry in the ground state breaks the degeneracy of the two AF magnon eigenmodes. We emphasize that although the spin Hamiltonian, Eq. (\ref{Hspin}), has $U(1)$ symmetry even in the easy-plane case, i.e., $\mathcal{K}_z = 0$, this symmetry is spontaneously broken in the ground state and the eigenmodes are not degenerate anymore \cite{PhysRevLett.118.137201}.  

To study spin excitations, we write the spin Hamiltonian, Eq. (\ref{Hspin}), in terms of so-called deviation bosonic operators using the Holstein-Primakoff (HP) transformation \cite{HolsteinPR1940}. We assume the ground state of the AF spin Hamiltonian is a collinear N{\'e}el state with ${\bf S}_{i \in \{{\cal A}\}}=(0,0,S)$ and ${\bf S}_{j \in \{{\cal B}\}}=(0,0,-S)$; and thus the HP transformation in the two-sublattice AF system reads,
\begin{subequations}
\begin{align}
&\label{HPa} S^z_i = S - a^\dag_i a_i ,\hspace{.25cm} 
S^-_i = a^\dag_i \sqrt{2S - a^\dag_i a_i},
\\
&\label{HPb}S^z_j = - S + b^\dag_j b_j ,\hspace{0.1cm} 
S^-_j = \sqrt{2S - b^\dag_j b_j} \; b_j,
\end{align}
\end{subequations}
and $S^+_{i(j)} = \big( S^-_{i(j)} \big)^\dag$, where $i$ and $j$ denote sites in the sublattices $\mathcal{A}$ and $\mathcal{B}$, respectively. The deviation operators $a$ ($b$) and $a^\dag$ ($b^\dag$) are, respectively, annihilation and creation operators at the sublattice $\mathcal{A}$ ($\mathcal{B}$) that follow the bosonic communication relations. The number of HP bosons at each site on sublattices $\mathcal{A}$ and $\mathcal{B}$ is upper-limited by $2S$ \cite{Konig}. Inserting Eqs. (\ref{HPa}) and (\ref{HPb}) into the AF spin Hamiltonian, Eq. (\ref{Hspin}), and under the assumption of small deviation number $\langle a^\dag_i a_i \rangle, \langle b^\dag_i b_i \rangle \ll 2S$, with $\langle\cdot\rangle$ the expectation value, we do a Taylor series expansion in the powers of $1/S$. This procedure leads to an effective bosonic Hamiltonian with an arbitrary interaction order,
\begin{align}\label{bosonH}
H_{b} = E^c_{0} + H_b^{(2)} + H_b^{(4)}+\cdots,
\end{align}
where ellipsis stands for higher order interactions.
The classical AF ground-state energy is given by,
\begin{align}\label{classic}
E^{c}_0=N S \left[-(z J/2 + \mathcal{K}_z) S + \mathcal{K}_x/2) \right],
\end{align}
where $z$ denotes the coordination number and $N$ is the total number of sites. $H_b^{(2)}$ and $H_b^{(4)}$ represent the noninteracting (quadric) and interacting (quartic) bosonic Hamiltonians, respectively.

The noninteracting Hamiltonian, consists of quadratic bosonic terms, in the momentum space, reads, 
\begin{align}\label{nonH}
\begin{split}
H^{(2)}_b =& S \sum_{\bs q} \bigg[ 
\big( z J + \mathcal{K}_x +  2 \mathcal{K}_z \big) \big( a^\dag_{\bs q} a_{\bs q} + b^\dag_{\bs q} b_{\bs q} \big) + zJ\gamma_{-{\bs q}}a_{\bs q} b_{-{\bs q}}
\\&
+ z J\gamma_{{\bs q}}b^\dag_{-{\bs q}} a^\dag_{\bs q}+\frac{\mathcal{K}_x}{2}\big( a_{\bs q} a_{-{\bs q}} + b_{\bs q} b_{-{\bs q}} + a^\dag_{\bs q} a^\dag_{-{\bs q}} + b^\dag_{\bs q} b^\dag_{-{\bs q}}  \big)
\bigg]
\end{split}
\end{align}
where $a_i = \sqrt{{2}/{N}} \sum_{\bs q} e^{i {\bs q} \cdot \mathbf{r}_i} a_{\bs q}$ and $b_j = \sqrt{{2}/{N}} \sum_{\bs q} e^{i {\bs q} \cdot \mathbf{r}_j} b_{\bs q}$, with $\{i,j\} \in \{\mathcal{A},\mathcal{B}\}$.
The lowest order interacting Hamiltonian, in our model, consists of quartic bosonic terms and describes the four-magnon scattering processes $H_b^{(4)}$. In section \eqref{sec:interaction}, we will investigate these nonlinear  terms in detail. In general, in AF spin systems, three-magnon interactions, consists of cubic bosonic terms, can be appeared in the presence of 
some spin interactions, like DMI and dipole-dipole interactions, as well as magnetic frustrations. These interactions break the conservation of total magnon numbers. However, in our model, Eq. \eqref{Hspin}, due to the absence of such spin interactions, three-magnon interactions are forbidden  \cite{ZhitomirskyRMP2013}.

\subsection{Linear spin-wave theory}
To investigate the effect of nonlinear four-magnon interactions on magnon dispersion and polarization, we should first find the band structure and eigenstates of the noninteracting magnon modes.
At a magnon density, we can disregard nonlinear magnon interactions. The qudratic bosonic Hamiltonian $H_b^{(2)}$ can be diagonalized via a standard canonical Bogolioubov transformation that maps the boson operators on two AF sublattices $\{a_{\bs q}, b_{\bs q}\}$, into the new bosonic magnon eigenmodes $\{\alpha_{\bs q}, \beta_{\bs q}\}$  where ${\bs q}$ is the magnon wavevector. In the rest of this section, we find the eigensystems of uniaxial and biaxial AF systems, separately.
\subsubsection{Uniaxial easy-axis AF system}\label{appAuni}
To diagonalize the noninteracting bosonic Hamiltonian (\ref{nonH}) in the uniaxial easy-axis limit ($\mathcal{K}_x = 0$), we perform a canonical transformation using the following $2\times 2$ Bogolioubov transformation, \cite{ShenPRB2019,RezendeJAP2019},
\begin{align}\label{BogoTuni}
    \begin{pmatrix}
    a_{\bs q} \\
    b^\dag_{-{\bs q}}
    \end{pmatrix}
    =
    \begin{pmatrix}
    \bar{u}^\mathrm{u}_{\bs q} & -\bar{v}^\mathrm{u}_{\bs q} \\
    -v^\mathrm{u}_{\bs q} & u^\mathrm{u}_{\bs q}
    \end{pmatrix}
    \begin{pmatrix}
    \alpha^\mathrm{u}_{\bs q} \\
    \beta^{\mathrm{u}\dag}_{\bs q}
    \end{pmatrix},
\end{align}
where $|u^\mathrm{u}_{\bs q}|^2 - |v^\mathrm{u}_{\bs q}|^2 = 1$.  We define $u^\mathrm{u}_{\bs q} = \cosh(\theta_{\bs q}/2)$ and $v^\mathrm{u}_{\bs q} = \sinh (\theta_{\bs q}/2)$. Inserting Eq. (\ref{BogoTuni}) into the Hamiltonian (\ref{nonH}), we obtain the following Bogoliubov coefficients for the uniaxial easy-axis system,
\begin{subequations}
\begin{align}\label{bogocoeffuni}
u^\mathrm{u}_{\bs q} &= \sqrt{\frac{z J S + 2 \mathcal{K}_z S + \epsilon^u_{\alpha,{\bs q}} }{2\epsilon^u_{\alpha,{\bs q}}}},
\\
v^\mathrm{u}_{\bs q} &= \sqrt{\frac{z J S + 2 \mathcal{K}_z S - \epsilon^u_{\beta,{\bs q}} }{2\epsilon^u_{\beta,{\bs q}}}}.
\end{align}
\end{subequations}
The degenerate eigenenergies of the circularly polarized eigenmodes $\chi\in\{\alpha,\beta\}$ of easy-axis AF system are,
\begin{equation}
    \epsilon^\mathrm{u}_{\chi,{\bs q}} = S\sqrt{\big( zJ + 2 \mathcal{K}_z
\big)^2-\big(z J |\gamma_{\bs q}|\big)^2},
\end{equation}
where $\gamma_{\bs q} = z^{-1}\sum_{i=1}^{z} e^{i {\bs q} \cdot \boldsymbol{\delta}_i}$ is the structure factor and $\boldsymbol{\delta}_i$ denotes the nearest-neighbor vectors.
\subsubsection{Biaxial easy-plane AF system}\label{appAbi}
In the case of biaxial AF systems, we use the following $4\times 4$ Bogolioubov transformations \cite{WhitePR1965, RezendeJAP2019},
\begin{equation}\label{bogoTbi}
\begin{split}
\begin{pmatrix}
a_{\bs q}
\\
b_{\bs q}
\\
a^\dag_{-{\bs q}}
\\
b^\dag_{-{\bs q}}
\end{pmatrix}    
=
\begin{pmatrix}
u_{\bs q} && v_{\bs q} && x_{\bs q} && w_{\bs q}
\\
-u_{\bs q} && v_{\bs q} && -x_{\bs q} && w_{\bs q}
\\
\bar{x}_{\bs q} && \bar{w}_{\bs q} && \bar{u}_{\bs q} && \bar{v}_{\bs q}
\\
-\bar{x}_{\bs q} && \bar{w}_{\bs q} && -\bar{u}_{\bs q} && \bar{v}_{\bs q}
\end{pmatrix}
\begin{pmatrix}
\alpha_{\bs q}
\\
\beta_{\bs q}
\\
\alpha^\dag_{-{\bs q}}
\\
\beta^\dag_{-{\bs q}}
\end{pmatrix}
\end{split}
\end{equation}
where the Bogoliubov coefficients are given by,
\begin{subequations}
\begin{align}
u_{\bs q}&\label{bogocoeffbi1}=\frac{1}{2}
\sqrt{\frac{ z J S + \mathcal{K}_x S +  2 \mathcal{K}_z S + \epsilon^\alpha_{{\bs q}}}{ \epsilon^\alpha_{{\bs q}}}}, 
\\
 v_{\bs q}&\label{bogocoeffbi2}=
- \frac{1}{2}\sqrt{\frac{ z J S + \mathcal{K}_x S +  2 \mathcal{K}_z S + \epsilon^\beta_{{\bs q}}}{\epsilon^\beta_{{\bs q}}}},
\\
 x_{\bs q}&\label{bogocoeffbi3}=
\frac{1}{2}\sqrt{\frac{ z J S + \mathcal{K}_x S +  2 \mathcal{K}_z S - \epsilon^\alpha_{{\bs q}}}{ \epsilon^\alpha_{{\bs q}}}},
\\
w_{\bs q}&\label{bogocoeffbi4}=\frac{1}{2}
\sqrt{\frac{ z J S + \mathcal{K}_x S +  2 \mathcal{K}_z S - \epsilon^\beta_{{\bs q}}}{ \epsilon^\beta_{{\bs q}}}}.
\end{align}
\end{subequations}
The eigenenergies of two magnon modes $\chi \in \{\alpha,\beta\}$ in easy-plane AF system are,
\begin{equation}\label{baredisp}
    \epsilon^{\chi}_{{\bs q}} = S \sqrt{\big( zJ+\mathcal{K}_x + 2 \mathcal{K}_z\big)^2-\big(z J |\gamma_{\bs q}| \mp \mathcal{K}_x\big)^2},
\end{equation}
where sign $-$($+$) refers to $\alpha$($\beta$) magnon mode.
The eignenenergies of the biaxial easy-plane case reduces to the uniaxial easy-axis limit when $\mathcal{K}_x = 0$, as it should be. \\ \\
In the easy-axis limit $\mathcal{K}_x=0$, and in the absence of external magnetic fields and Dzyaloshinskii-Moriya interactions (DMIs), the two magnon eigenenergies are degenerate and the eigenmodes are circularly polarized with opposite chiralities \cite{RezendeJAP2019,PhysRevB.97.020402}.
The presence of a hard-axis anisotropy ($\mathcal{K}_x\neq 0$) removes the degeneracy of magnon eigenstates and causes the two eigenmodes become elliptically polarized \cite{RezendeJAP2019,PhysRevB.97.020402}. The latter is a consequence of the breaking of $U(1)$ symmetry in the ground state of the spin Hamiltonian \ref{Hspin}. In the presence of the long-range dipolar interactions that also breaks U(1) symmetry, the magnon eigenmodes of a uniaxial easy-axis are elliptically polarized \cite{PhysRevLett.124.077201}. In the present study, we ignore the effect of dipolar interaction since they are negligible in most of AF systems. \\

After diagonalization, the total noninteracting AF Hamiltonian in the new eigenbasis is give by,
\begin{align}\label{Hm}
H_m&=E_0+{H}_m^{(2)},\\
E_0&=E^c_0 + E^q_0,
\end{align}
where $E_0$ is the total AF ground-state energy,
and the quantum correction to the classic AF ground-state energy, $E^c_0$, reads,
\begin{align}\label{E0quantum}
\begin{split}
E_0^q 
=
S \sum_{\bf q}
\Big\{ &
2 (z J + \mathcal{K}_x + 2 \mathcal{K}_z) (|x_{\bf q}|^2+|w_{\bf q}|^2)
\\&
+
2 \, \text{Re} \big[ z J \gamma_{-\bf q} (- u_{\bf q} x_{\bf q} + v_{\bf q} w_{\bf q} )
+
\mathcal{K}_x (u_{\bf q} x_{\bf q} + v_{\bf q} w_{\bf q} )
\big]\Big\}.
\end{split}
\end{align}
This quantum zero-point energy lowers the classical ground-state energy and expresses the well-known fact that the true ground state of an AF system is not a N{\'e}el state \cite{QuantumMagnetism}.   
The quadratic part of the AF Hamiltonian that describes the dynamics of magnons is,
\begin{align}\label{noninteracting}
{H}_m^{(2)} = \sum_{\bs q} \left[\epsilon^{\alpha}_{\bs q }\alpha^\dag_{\bs q} \alpha_{\bs q}+\epsilon^{\beta}_{\bs q}\beta^\dag_{\bs q} \beta_{\bs q}\right].
\end{align}
where the bare eigenenergy for the magnon mode $\chi\in \{\alpha,\beta\}$ is given by Eq.~\eqref{baredisp}. 

The structure factors for 2D square ($z=4$) and hexagonal ($z=3$) lattices are, $\gamma_{\bs q}=2\left(\cos(q_x/q_m)+\cos(q_y/q_m)\right)/z$ and $\gamma_{\bs q}=e^{iq_x/q_m} \left(1+2 e^{-i 3 q_x/2q_m}\cos(\sqrt{3}q_y/2q_m)\right)/z$, respectively, where $q_m=1/a$ and $a$ is the lattice constant. 
In the long-wavelength limit ${\bm q} \rightarrow 0$, the structure factors of both hexagonal and square lattices reduce to $\gamma_{\bs q} \simeq 1-q^2/4q_m^2$; and thus the mode-dependent eigenenergies becomes to $\epsilon^{\alpha}_\mathbf{q} =S \sqrt{ 4(\mathcal{K}_z+\mathcal{K}_x)(z J+\mathcal{K}_z)+z J q^2(z J - \mathcal{K}_x)/2q_m^2}$ and $\epsilon^{\beta}_{\mathbf{q}}= S \sqrt{4\mathcal{K}_z(z J + \mathcal{K}_z+\mathcal{K}_x)+z J q^2(z J + \mathcal{K}_x)/2q_m^2}$. In other words, around the center of the Brillouin zone, called $\Gamma$-point, the two low-energy AF magnon modes propagate as relativistic-like quasiparticles with different effective \emph{rest masses} and \emph{speed of light}, determined by the magnetic exchange stiffness and magnetic anisotropies. 
The bare magnon bandgaps of each mode are $\epsilon^{\alpha}_{\mathbf{q}=0}$ and $\epsilon^{\beta}_{\mathbf{q}= 0}$ that can be measured by in AF resonance experiments. The presence of this gap stabilizes long-range magnetic order at finite temperature in 2D magnetic systems and the magnetic transition temperatures in these systems are proportional to this energy gap. 
We will later show how the bare magnon dispersions become renormalized in the presence of nonlinear thermal magnon interactions. 

\section{Mean-Field theory of Magnon-Magnon Interactions} \label{sec:interaction}
In this section, we use a Hartree-Fock mean-field theory to treat the nonlinear thermal magnon interactions in a two-sublattice AF system at finite temperature. The lowest-order nonlinear terms related to two-boson scattering are represented by four-boson interactions,
\begin{widetext}
\begin{align}\label{eq:Hintdelta}
H^{(4)}_b=\frac{-J}{4}\sum_{\langle i, j \rangle} \Big[ a^\dag_i a_i b^\dag_j b_j + a_i b^\dag_j b_j b_j + (a\leftrightarrow b)\Big]
-\frac{\mathcal{K}_x}{8} \sum_{i}\Big[
\Delta\mathcal{K}a^\dag_i a^\dag_i a_i  a_i+ a^\dag_i a_i a_i a_i+(a\leftrightarrow b)\Big]+\text{H.c.},
\end{align}
\end{widetext}
with $\Delta\mathcal{K}=2(1+\mathcal{K}_z/\mathcal{K}_x)$, and \text{H.c.} is the Hermitian conjugate.
Four-boson interactions in AF systems are more complicated than the ferromagnetic one.  In ferromagnets, in the absence of long-range dipolar fields, the four-boson interactions conserve the number of quasiparticles because there are only quartic terms with two creation and two annihilation operators. On the contrary, as it can be seen in Eq. (\ref{eq:Hintdelta}), in AF systems the magnon number could in general be a nonconserved quantity. It was also shown that spontaneous three-magnon and in general all other $n$-magnon decays are forbidden in collinear AF systems by energy conservation at zero temperature in the absence of an external magnetic field  \cite{HarrisPRB1971, ZhitomirskyRMP2013}.

To explore the effects of nonlinear magnon interactions on the bare spectra of magnons at finite temperature, we implement a mean-field theory based on the self-consistent Hartree-Fock approximation \cite{balucani1980magnetic,OguchiPR1960} to reduce Eq. (\ref{eq:Hintdelta}) into an effective noninteracting Hamiltonian in terms of quadratic magnon operators. In the mean-field treatment, we start out by substituting
$\xi^\dag_i{\xi}'_j=\langle\xi^\dag_i{\xi}'_j\rangle_{\text{th}}+\Xi_{\xi{\xi}'}$ into the Hamiltonian \ref{eq:Hintdelta}, where $\xi_i,{\xi}'_i\in\{a_i,b_i,a_i^{\dagger},b_i^{\dagger}\}$ and $\langle\cdot\rangle_{\text{th}}$ is a thermal average. We expand up to the first order in the field $\Xi_{\xi{\xi}'}\equiv\xi^\dag_i{\xi}'_j-\langle\xi^\dag_i{\xi}'_j\rangle_{\text{th}}$, and express the operators in the basis of $\alpha$- and $\beta$-magnons through the Bogoliubov transformation. In the leading-order correction, it is assumed that mean-field parameter $\Xi_{\xi{\xi}'}$ characterizes slight nonequilibrium deviations from the thermal equilibrium, which is consistent with our earlier assumption of small deviation in the boson number in the Taylor series expansion of HP transformations. To guarantee the thermalization of each magnon mode, we assume that nonlinear magnon scattering-times are much faster than the timescales associated with spin non-conserving interactions such as magnon-phonon interactions \cite{TroncosoPRB2020}. In the Hartree-Fock approximation, we only keep the diagonal terms and assume for each bosonic operator $\chi_{\bs q}\in \{\alpha_{\bs q},\beta_{\bs q}\}$, so that $\langle\chi^{\dagger}_{\bs q}\chi_{\bs q'}\rangle_{\text{th}}=\delta_{{\bs q},{\bs q}'}n_{\chi,{\bs q}}$ \cite{LI201828,PhysRevB.104.064435}. The thermodynamic equilibrium of the AF magnon number at each magnon mode $\chi\in \{\alpha,\beta\}$ is defined through the Bose-Einstein distribution function $n_{\chi,{\bs q}}=(e^{\epsilon_{\chi,{\bs q}}/k_B T} - 1)^{-1}$, with the Boltzmann constant $k_B$, and the temperature $T$. The chemical potential of magnons is zero at thermal equilibrium \cite{FlebusPRB2019}.

The effective mean-field Hamiltonian in momentum space and first order in the fluctuations $\Xi_{\xi{\xi}'}$, becomes ${H}_{\text{MF}}^{(4)}=\frac{1}{2}\sum_{\bs q}\Psi^{\dagger}_{\bs q}\mathbb{H}^{(4)}\Psi_{\bs q}$,
with the vector operator $\Psi_{\bs q}=(\alpha_{\bs q},\beta_{\bs q},\alpha^{\dagger}_{-\bs q},\beta^{\dagger}_{-\bs q})^{T}$. The matrix Hamiltonian $\mathbb{H}^{(4)}$ is defined by,
\begin{align}\label{eq:matrixH}
\mathbb{H}^{(4)}=\left(
\begin{array}{cccc}
{\cal C}_{\alpha\alpha} & {\cal D}_{\alpha\beta} & 2\bar{\cal C}'_{\alpha\alpha} & \bar{\cal D}'_{\alpha\beta}\\
\bar{\cal D}_{\alpha\beta} & {\cal C}_{\beta\beta} & \bar{\cal D}'_{\alpha\beta} & 2\bar{\cal C}'_{\beta\beta}\\
2{\cal C}'_{\alpha\alpha} & {\cal D}'_{\alpha\beta} & {\cal C}_{\alpha\alpha} & \bar{\cal D}_{\alpha\beta}\\
{\cal D}'_{\alpha\beta} & 2{\cal C}'_{\beta\beta} & {\cal D}_{\alpha\beta} & {\cal C}_{\beta\beta}
\end{array}\right),
\end{align} 
where the overbar represents the complex conjugate and the explicit dependence of the temperature-dependent mean-field coefficients on the wavevector and temperature is omitted for brevity. The Hamiltonian ${H}_{\text{MF}}^{(4)}$ represents the quantum and temperature-dependent corrections to the noninteracting magnon Hamiltonian, when the lowest-order nonlinear AF dynamics is taken into account. The coefficients of the matrix $\mathbb{H}^{(4)}$ depend on temperature, for more details see Appendix \ref{appBbi}. This effective Hamiltonian ${H}_{\text{MF}}^{(4)}$ describes scattering processes between magnons of the same band (intraband contribution) and magnons of different bands (interband contribution). The components ${\cal C}_{\chi\chi}$ and ${\cal C}'_{\chi\chi}$, describing intraband processes of magnons at the $\chi$-band, are given by Eqs.~(\ref{simplecoeffHbi}) in Appendix~(\ref{appBbi}).
\begin{subequations}\label{intracoeffHbi}
\begin{align}
{\cal C}_{\alpha\alpha,{\bs q}} & = \Delta^{-}_{\bs q} |u_{\bs q}|^2 + \Delta^{-}_{-\bs q}|x_{\bs q}|^2 + 2 \, \text{Re} \left[ u_{\bs q} x_{\bs q} \, \big(\xi - \text{Re} \, \Gamma_q  \big) \right],
\\
{\cal C}_{\beta\beta,{\bs q}} &= \Delta^{+}_{\bs q} |v_{\bs q}|^2 + \Delta^{+}_{\bs q} |w_{\bs q}|^2 +  2 \, \text{Re} \left[ v_{\bs q} w_{\bs q} \, \big(\xi + \text{Re} \, \Gamma_q  \big) \right],
\\
{\cal C}^\prime_{\alpha\alpha,{\bs q}}&=2\Delta^{-}_{\bs q}u_{\bs q}{x}_{\bs q}+\Lambda^{-}_{\bs q} u^2_{\bs q}+ \bar{\Lambda}^{-}_{\bs q}{x}^2_{\bs q},
\\
{\cal C}^\prime_{\beta\beta,{\bs q}}&=2\Delta^{+}_{\bs q}v_{\bs q}{w}_{\bs q}+ \Lambda^{+}_{\bs q} v^2_{\bs q}+ \bar{\Lambda}^{+}_{\bs q} {w}^2_{\bs q},
\end{align}
\end{subequations}
where $\Delta^{\pm}_{\bs q}$, $\xi$, $\Gamma_{\bs q}$ and $\Lambda^{\pm}_{\bs q}$ are defined in Eq.~(\ref{simpcoeff}). These mean-field coefficients parameterize the amplitude of the intra- and inter-band magnon interactions. The relevant Bogoliubov coefficients  $u_{\bs q},v_{\bs q},x_{\bs q}$ and $w_{\bs q}$ are defined by Eqs. (\ref{bogocoeffbi1})-(\ref{bogocoeffbi4}). Interband scattering of magnons are characterized by the components ${\cal D}_{\alpha\beta}$ and ${\cal D}^\prime_{\alpha\beta}$, and read,
\begin{subequations}\label{intercoeffHbi}
\begin{align}
{\cal D}_{\alpha\beta,{\bs q}}=&\Gamma_{\bs q}x_{\bs q}v_{\bs q}-\bar{\Gamma}_{\bs q}{u}_{\bs q}{w}_{\bs q}+\Theta_{-\bs q}x_{\bs q}{w}_{\bs q}-\Theta_{\bs q}v_{\bs q}{u}_{\bs q}
 \nonumber\\&
+ 2 \, \xi v_{\bs q} \, x_{\bs q} + 2 \, \bar{\xi} \, u_{\bs q} \, w_{\bs q},
\\
{\cal D}^\prime_{\alpha\beta,{\bs q}}=& \bar{\Gamma}_{\bs q}{x}_{\bs q}{w}_{\bs q} - \Gamma_{\bs q} u_{\bs q} v_{\bs q} - {\Theta}_{-\bs q}{x}_{\bs q} v_{\bs q} + \Theta_{\bs q}{w}_{\bs q} u_{\bs q}
 \nonumber\\&
+ 2 \xi^\prime u_{\bs q} v_{\bs q} + 2 \, \bar{\xi}^\prime \, x_{\bs q} w_{\bs q},
\end{align}
\end{subequations}
where $\Gamma^\prime_{\bs q}$, $\Theta_{\bs q}$ and $\xi^\prime$ can be found in Eqs. (\ref{simpcoeff}). Through the interband process, $\alpha$- and $\beta$-magnon modes become coupled and thus, hybridized. The definition of $A_a,\, A^\prime_a,\, A_{ab,{\bs q}}$ and $A^\prime_{ab,{\bs q}}$ are respectively detailed in Eqs. (\ref{eq:A'a})-(\ref{eq:A'ab}) in the Appendix \ref{appB}.  The mean-field coefficients ${\cal C}_{\chi\chi},\,{\cal C}'_{\chi\chi},\,{\cal D}_{\alpha\beta}$ and ${\cal D}'_{\alpha\beta}$ are functions of temperature and magnon energy through the Bose-Einstein distribution $n_{\chi,\bs q}$. The evaluation of these coefficients, as well as to obtain the eigenenergies, requires a self-consistent calculation. 

The analytical expressions obtained for the mean-field coefficients, Eqs. (\ref{intracoeffHbi})-(\ref{intercoeffHbi}) constitute the main result of our work. To illustrate the effects of interacting thermal magnons on the AF magnon spectrum and the polarization of magnons, we analyze both uniaxial (${\cal K}_x=0$) and biaxial (${\cal K}_x\neq 0$) AF systems with hexagonal and square lattice structures that are two common lattice geometries.

\begin{figure}
    \includegraphics[width=\columnwidth]{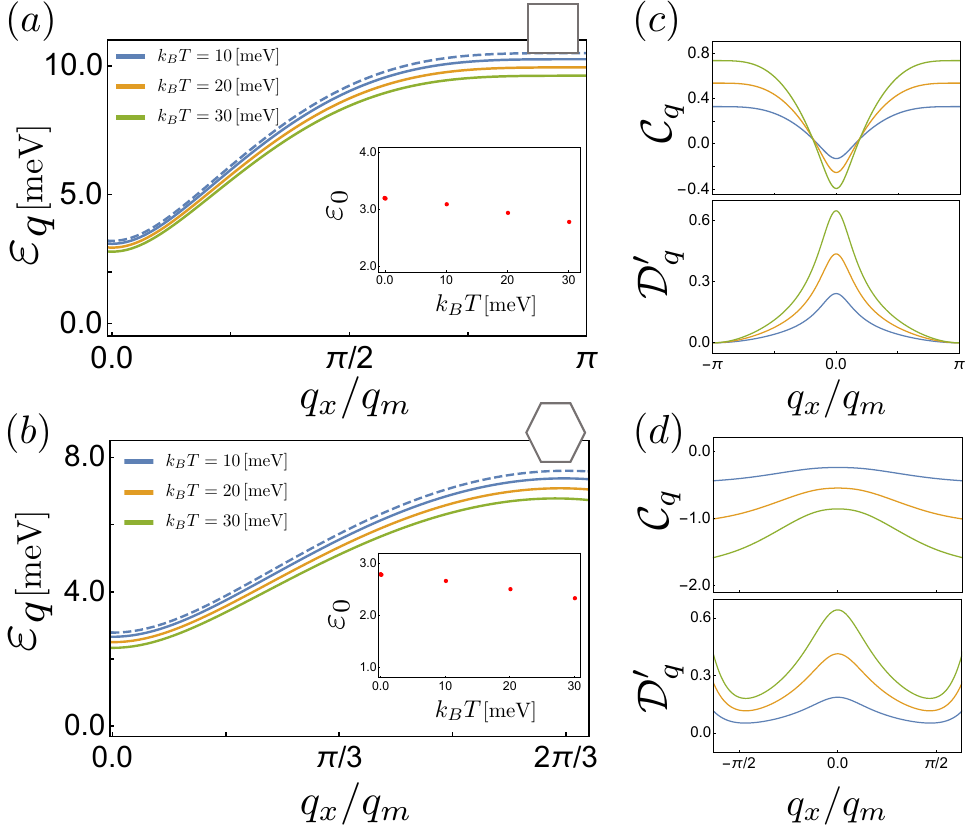}
    \caption{Temperature dependence of degenerate magnon eigenenergies of a uniaxial AF system with (a) 2D square and (b) 2D hexagonal lattices as a function of dimensionless wavevector $q_x/q_m$. Dashed curve lines represent the bare magnon dispersion given by Eq. (\ref{baredisp}) with $\mathcal{K}_x = 0$. At the inset of panels (a) and (b), the AF resonance modes (renormalized magnon bandgap) as a function of temperature are presented. The mean-field coefficients (in units of meV) , ${\cal C}$ and ${\cal D}'$,  are shown for different temperatures as a function of wavevector for square and hexagonal lattices at panels (c) and (d), respectively. We use these typical parameters: $J = 1$ [meV], $\mathcal{K}_z = 0.1$ [meV], $S=5/2$ and $a=1$ [nm].}
    \label{fig:uni-meanfield}
\end{figure}

\section{Nonlinear spectral shift of magnons} \label{renormalization}
In this section, we investigate the effect of nonlinear magnon interactions on magnon dispersion, magnon bandgaps, and AF resonance modes of easy-axis and easy-plane antiferromagnetic insulators.

\subsection{Uniaxial easy-axis AF systems}\label{interactionuni}
First, for the sake of completeness, we consider an AF system with easy-axis anisotropy along the $z$-direction, modeled by the Hamiltonian in Eq. (\ref{Hspin}) when $\mathcal{K}_x=0$. Nonlinear magnon interactions in uniaxial AF systems have been previously treated in detail to unveil their role in the lifetime of magnons \cite{HarrisPRB1971,BayrakciPRL2013}, hydrodynamic regime \cite{HarrisPRB1971} and bulk spin transport \cite{ShenPRB2019,TroncosoPRB2020}.

As we have already mentioned, in the uniaxial AF system and in the absence of a magnetic field and DMIs, the two AF magnon eigenmodes are degenerate and given by Eq. (\ref{baredisp}) with $\mathcal{K}_x=0$, and their corresponding eigenstates are circularly polarized magnons. On the other hand, the effective interacting Hamiltonian of magnons in the uniaxial AF case, in a basis that diagonalizes the bare part and within the mean-field approximation, see Appendix \ref{appAuni}; is given by,
\begin{align}\label{effectivH4uni}
{H}^{\mathrm{u}(4)}_{\text{MF}}=\sum_{\bs q}\left[{\cal C}_{\bs q}
\alpha^{\mathrm{u}\dagger}_{\bs q}\alpha^\mathrm{u}_{\bs q} + {\cal C}^\prime_{\bs q} \beta^{\mathrm{u}\dag}_{\bs q}\beta^\mathrm{u}_{\bs q} + {\cal D}'_{\bs q}\alpha^\mathrm{u}_{\bs q}\beta^\mathrm{u}_{-\bs q}+\text{H.c.}
\right],
\end{align}
where the superscript $"\mathrm{u}"$ refers to the uniaxial case. The mean-field coefficients ${\cal C}_{\bs q}$ and ${\cal C}^\prime_{\bs q}$ renormalize two noninteracting magnon modes, described by the linear spin-wave theory $H^{(2)}_{\text{m}}$, Eq. (\ref{noninteracting}). In the absence of magnetic fields and DMIs ${\cal C}_{\bs q}={\cal C}^\prime_{\bs q}$.

The magnon nonconserving term with strength ${\cal D}^\prime_{\bs q}$, that represents interband magnon scatterings with opposite linear momenta, breaks the local $U(1)$ symmetry associated with the magnon number conservation. 
This process is equivalent to the excitation and annihilation of a magnon pair with zero net linear and spin angular momenta.  We later show how this term leads to the thermal squeezing of magnon eigenmodes.

The temperature-dependent mean-field coefficients are given by,
\begin{subequations}
\begin{align}
{\cal C}_{\bs q}&\label{eq:unimeanH-C} = \frac{1}{2} [A_1 (|u^\mathrm{u}_{\bs q}|^2 + |v^\mathrm{u}_{\bs q}|^2)- (A_{2, \bs q} \,  \bar{u}^\mathrm{u}_{\bs q} \bar{v}^\mathrm{u}_{\bs q} + \bar{A}_{2, \bs q} \, u^\mathrm{u}_{\bs q} v^\mathrm{u}_{\bs q})],
\\
{\cal C}^\prime_{\bs q}& \label{eq:unimeanH-C'}= \frac{1}{2} [A_1 (|u^\mathrm{u}_{\bs q}|^2 + |v^\mathrm{u}_{\bs q}|^2)- (A_{2, -\bs q} \, \bar{u}^\mathrm{u}_{\bs q} \bar{v}^\mathrm{u}_{\bs q} + \bar{A}_{2, -\bs q} \, u^\mathrm{u}_{\bs q} v^\mathrm{u}_{\bs q})],
\\
{\cal D}^\prime_{\bs q}&\label{eq:unimeanH-D'} =
-2 A_1 \bar{u}^\mathrm{u}_{\bs q} v^\mathrm{u}_{\bs q} + [\bar{A}_{2, \bs q} (v^{\mathrm{u}}_{\bs q})^{2} + A_{2,\bs q} (\bar{u}^{\mathrm{u}}_{\bs q})^2],
\end{align}
\end{subequations}
the Bogoliubov parameters of the uniaxial case, $u^\mathrm{u}_{\bs q}$ and $v^\mathrm{u}_{\bs q}$, and temperature-dependent mean-field coefficients $A_1$ and $A_2$ are presented in Appendix \ref{appB}. Even at zero temperature, these coefficients are finite. Therefore, we can generally say that there is always a finite {\it{nonlinear}} quantum correction to the bare magnon dispersion in AF systems.  

The resulting total magnon Hamiltonian of a uniaxial AF system in the mean-field approximation is $H^\mathrm{u}={H}^{\mathrm{u}(2)}_{\text{m}}+{H}^{\mathrm{u}(4)}_{\text{MF}}$. Due to the presence of interband processes, ${\cal D}^\prime$, $\alpha$ and $\beta$ modes are coupled and the Hamiltonian $H^\mathrm{u}$ is no longer diagonal in the $\{\alpha,\beta\}$ basis. To diagonalize $H^\mathrm{u}$, we should introduce a new diagonal basis of operators, $\hat{\mu}_{\bs q}$ and $\hat{\nu}_{\bs q}$, and the Hamiltonian in this new basis reads $H^\mathrm{u}=\sum_{\bs q}\left(\varepsilon^{\mu}_{\bs q}\hat{\mu}^{\dagger}_{\bs q}\hat{\mu}_{\bs q}+\varepsilon^{\nu}_{\bs q}\hat{\nu}^{\dagger}_{\bs q}\hat{\nu}_{\bs q}\right)$. In the absence of magnetic fields and DMIs, the effective eigenmodes of an interacting uniaxial AF system remains degenerate, $\varepsilon^{\mu}_{\bs q}=\varepsilon^{\nu}_{\bs q}\equiv\varepsilon_{\bs q}$, where
\begin{align}\label{eq.eigen-ene-uni}
\varepsilon_{\bs q}=\sqrt{\left(\epsilon^{\text{u}}_{\bs q}+{\cal C}_{\bs q}\right)^2-|{\cal D}'_{{\bs q}}|^2},
\end{align}
with $\epsilon^{\text{u}}_{\bs q}=z JS\sqrt{
\big(1+2\mathcal{K}_z/zJ
\big)^2-|\gamma_{\bs q}|^2}$ denotes the dispersion relation of noninteracting magnons in the uniaxial case, i.e., Eq. (\ref{baredisp}) with $\mathcal{K}_x = 0$.

The dispersion in Eq. (\ref{eq.eigen-ene-uni}) is valid for two-sublattice AF systems with arbitrary dimensions and lattice structures at finite temperature. Here, we consider two distinct geometries of spin systems which are quite common, i.e., square and hexagonal lattices. The evaluation of mean-field coefficients, Eqs. (\ref{eq:unimeanH-C})-(\ref{eq:unimeanH-D'}), and the eigenenergy $\varepsilon_{\bs q}$, is done self-consistently and involves the calculation of $A_1$ and $A_2$ for various temperatures.

In Fig. \ref{fig:uni-meanfield}, we display the magnon eigenenergies and mean-field parameters as a function of the wavevector for different temperatures and for two lattice structures. As we have already discussed, the eigenenergies do not reduce to the eigenenergy of noninteracting magnons $\epsilon^{\text{u}}_{\bs q}$ at zero temperature (dashed line at Fig. \ref{fig:uni-meanfield}(a) and (b)),  thanks to the existence of nonlinear quantum corrections. At finite temperature, we observe that magnon interactions lower the energy of long wavelength magnons. In particular, from the inset of panels (a) and (b), we see that AF resonance modes are decreased by increasing the temperature. This figure shows that nonlinear magnon interactions are more pronounced in AF systems with hexagonal lattice than their square lattice counterparts. 

The mean-field parameters ${\cal C}_{\bs q}$ and ${\cal D}'_{\bs q}$, plotted at Fig. \ref{fig:uni-meanfield}(c) and (d) for each lattice, are both peaked at the zero wavevector. The intraband coefficient ${\cal C}_{\bs q}$ may change the sign by varying the temperature and wavevector while the interband coefficient ${\cal D}'_{\bs q}$ is always positive in our formalism.

\begin{figure}
    \includegraphics[width=\columnwidth]{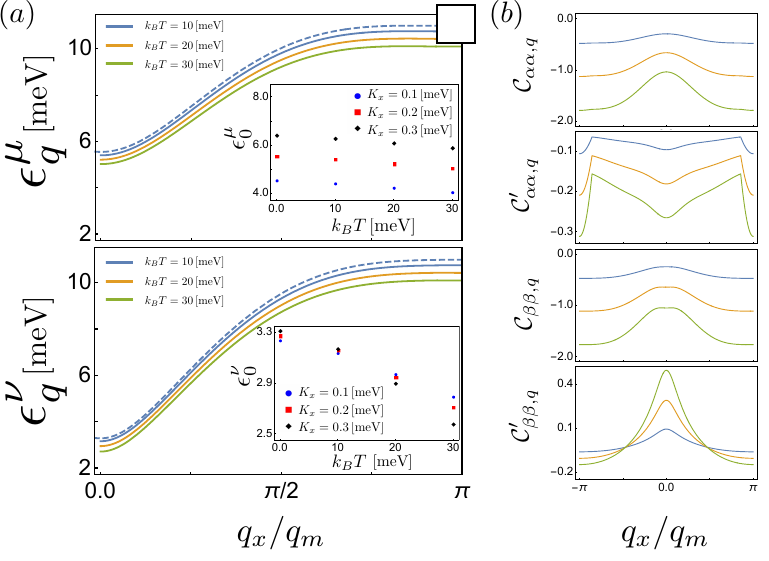}
    \caption{Temperature dependence of magnon eigenenergies of a biaxial AF system in a 2D square lattice. In panel (a), the eigenenergies for both $\mu$ and $\nu$ modes as a function of the wavevector are plotted. Dashed lines represent the noninteracting eigenenergies, $\epsilon^{\alpha}_{\bs q}$ and $\epsilon^{\beta}_{\bs q}$, see Eq. (\ref{baredisp}). The AF resonance modes, $\epsilon^{\mu}_{ 0}$ and $\epsilon^{\nu}_{ 0}$, are shown in their respective panels, as a function of temperature and different hard-axis anisotropy strengths. In panel (b), the mean-field coefficients (in meV units) ${\cal C}_{\alpha\alpha},{\cal C}'_{\alpha\alpha}, {\cal C}_{\beta\beta}$ and ${\cal C}'_{\beta\beta}$, are plotted for various temperatures as a function of wavevector. The parameters employed are $J = 1$ [meV], $\mathcal{K}_x=2\mathcal{K}_z = 0.2$ [meV], $S=2.5$ and $a=1$ [nm].}
    \label{fig:biaxial-meanfield}
\end{figure}

\subsection{Biaxial easy-plane AF systems}\label{interactionbi}
In this subsection, we focus on the AF systems with biaxial anisotropy. Previously, we have shown that the degeneracy of two noninteracting magnon modes is lifted in this case, see Eq. (\ref{baredisp}), and the corresponding eigenstates are elliptically polarized magnons \cite{RezendeJAP2019}.
The total interacting magnon Hamiltonian within the mean-field approximation is $H=H^{(2)}_{\text{m}}+H^{(4)}_{\text{MF}}$. The mean-field Hamiltonian (\ref{eq:matrixH}) can be partitioned into the intra- and interband contribution as $H^{(4)}_{\text{MF}}=H^{(4)}_{\text{intra}}+H^{(4)}_{\text{inter}}$, with
\begin{subequations}\label{effectivH4bi}
\begin{align}
{H}_{\text{intra}}^{(4)}\nonumber=&\label{effectivH4intra}\sum_{\bs q}\Bigl[{\cal C}_{\alpha\alpha,{\bs q}}\alpha^\dag_{\bs q} \alpha_{\bs q}+{\cal C}_{\beta\beta,{\bs q}}\beta^\dag_{\bs q}\beta_{\bs q}\\
&
+{\cal C}'_{\alpha\alpha,{\bs q}}\alpha_{\bs q}\alpha_{-{\bs q}}
+{\cal C}'_{\beta\beta,{\bs q}}\beta_{\bs q} \beta_{-{\bs q}}+\text{H.c.}\Bigr],\\
{H}_{\text{inter}}^{(4)}=&\label{effectivH4inter}\sum_{\bs q}\left[{\cal D}_{\alpha\beta,{\bs q}}\alpha^\dag_{\bs q}\beta_{\bs q}+{\cal D}'_{\alpha\beta,{\bs q}}\alpha_{\bs q}\beta_{-{\bs q}}+\text{H.c.}\right].
\end{align}
\end{subequations}
The intraband contribution of nonlinear magnon scattering consists of diagonal elements, in the $\alpha$ and $\beta$ operators representation, with coefficients ${\cal C}_{\chi\chi}$ and the off-diagonal elements with coefficients ${\cal C}'_{\chi\chi}$. ${\cal C}_{\chi\chi=\alpha\alpha(\beta\beta)}$ renormalizes noninteracting magnon dispersions, obtained within the linear spin-wave theory $H^{(2)}_{\text{m}}$, Eq. (\ref{noninteracting}), while ${\cal C}'_{\chi\chi=\alpha\alpha(\beta\beta)}$ characterize magnon scatterings between states with opposite momenta, $\bm{q}$ and $-\bm{q}$, inside a particular magnon band. The later is equivalent to the annihilation and excitation of an (a) $\alpha (\beta)-$magnon pair with zero net linear momentum and spin angular momentum $+2\hbar (-2\hbar)$.
The interband nonlinear magnon scatterings are parameterized by off-diagonal terms, ${\cal D}_{\alpha\beta}$ and ${\cal D}'_{\alpha\beta}$, corresponding to scattering processes of magnons between $\alpha$ and $\beta$ bands. The total interacting Hamiltonian, $H$, is no longer diagonalized in the basis of $\alpha$ and $\beta$ operators. Thus, we do another canonical transformation and find the diagonal basis in terms of new bosonic operators, say $\mu$ and $\nu$.
The expressions for mean-field parameters are presented in the Appendix \ref{appBbi}, see Eqs. (\ref{eq:meanfield-m})-(\ref{eq:meanfield-p'}). In the biaxial case, similar to the uniaxial one, the temperature-dependent mean-field coefficients are finite even at $T=0$. This {\it{nonlinear}} quantum corrections to magnon dispersion scales as $1/N$, see the Appendix \ref{appBbi}, and therefore are negligible for large systems. The nonlinear quantum corrections, in addition to linear quantum spin fluctuations, could play a relevant role in the quantum spin-transfer torques and spin shot noise phenomena \cite{QaiumzadehPRB2018}. The importance of this correction is, however, conditioned to small systems, which is a size limit where our mean-field approach might breakdown.
\begin{figure}[tbh]
    \includegraphics[width=\columnwidth]{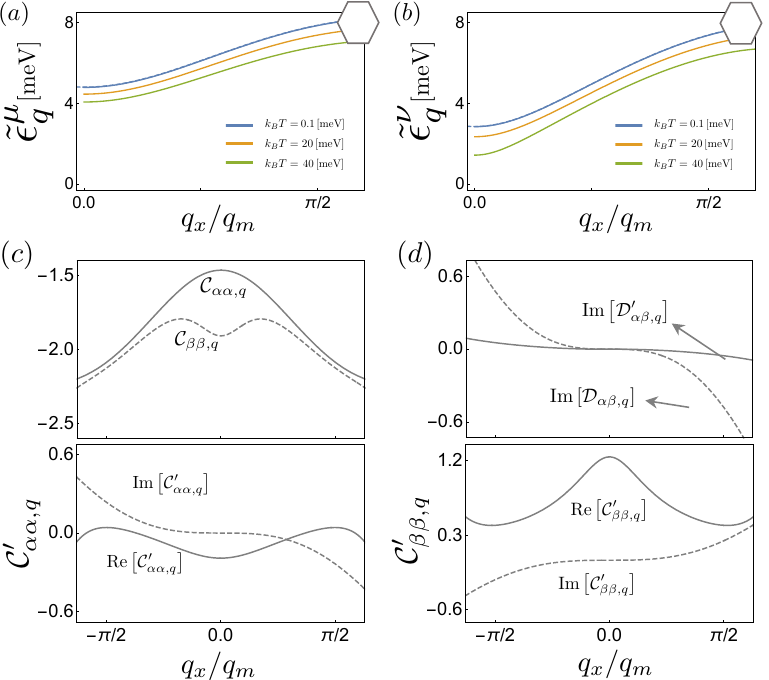}
    \caption{Temperature dependence of magnon eigenenergies of a biaxial AF system in a 2D hexagonal lattice. In panels (a) and (b), the energies $\tilde{\epsilon}^{\mu }_{q}$ and $\tilde{\epsilon}^{\nu }_{q}$ are plotted as a function of the wavevector. Dashed lines represent the respective noninteracting energies, see Eq. (\ref{baredisp}). The mean-field coefficients (in meV units) ${\cal C}_{\alpha\alpha}, {\cal C}'_{\alpha\alpha}, {\cal C}_{\beta\beta}, {\cal C}'_{\beta\beta}, {\cal D}_{\alpha\beta}$ and ${\cal D}'_{\alpha\beta}$, are shown as a function of the wavevector in panels (c) and (d) with $k_BT=40 [\text{meV}]$ and $K_x=0.2 [\text{meV}]$. Note that the real parts of coefficients ${\cal D}_{\alpha\beta}$ and ${\cal D}'_{\alpha\beta}$ are zero, and thus only imaginary parts are displayed. The parameters employed are the same as in Fig. \ref{fig:biaxial-meanfield}.}
    \label{fig:biaxial-meanfield-hexagonal}
\end{figure}

We find that in square lattices ${\cal D}_{\alpha\beta}={\cal D}'_{\alpha\beta}=0$, hence, the interband scattering processes are absent. The eigenenergies for the two magnon bands in a square lattice are given by,
\begin{subequations}\label{E-sq}
\begin{align}
\epsilon^{\mu}_{\bs q}&=\sqrt{(\epsilon^{\alpha}_{{\bs q}}+{\cal C}_{\alpha\alpha,{\bs q}})^2-|{\cal C}'_{\alpha\alpha,{\bs q}}|^2},
\\
\epsilon^{\nu}_{\bs q}&=\sqrt{(\epsilon^{\beta}_{{\bs q}}+{\cal C}_{\beta\beta,{\bs q}})^2-|{\cal C}'_{\beta\beta,{\bs q}}|^2}.
\end{align}
\end{subequations}
In Fig. \ref{fig:biaxial-meanfield}(a), we show the magnon eigenenergy of a 2D square lattice as a function of the wavevector for different temperatures. The situation is similar in 3D AF systems, see the Appendix \ref{app:eigenenergiesbiaxial}. The noninteracting eigenenergies, $\epsilon^{\alpha}_{\bs q}$ and $\epsilon^{\beta}_{\bs q}$, in the upper- and lower-panel are respectively represented by dashed lines. The effect of nonlinear interactions is in particular evidenced by the temperature dependence of the AF resonance modes, $\epsilon^{\mu}_{ 0}$ and $\epsilon^{\nu}_{ 0}$, shown in the respective insets of panel (a). These figures show that the energy of one magnon band $\epsilon^{\nu}_{0}$, decreases with temperature, while another band $\epsilon^{\mu}_{ 0}$ has a weak temperature dependence for different strengths of the hard-axis anisotropy.

Quite contrary, we find that in hexagonal lattices, the interband mean-field coefficients ${\cal D}_{\alpha\beta}$ and ${\cal D}^\prime_{\alpha\beta}$ are nonzero. The  eigenenergies for the magnonic $\mu$- and $\nu$-eigenmodes, denoted by $\tilde{\epsilon}^{\mu}_{\bs q}$ and $\tilde{\epsilon}^{\nu}_{\bs q}$, respectively are shown in Fig. \ref{fig:biaxial-meanfield-hexagonal}. Their analytical expressions are displayed in Eq. (\ref{eq:magnonenergies-biaxial}) at Appendix \ref{app:eigenenergiesbiaxial}.

Generally, in biaxial cases, the degeneracy of two magnon bands are broken and one band (the $\nu$ band) has lower energy than other one (the $\mu$ band). Therefore, increasing temperature leads to an increase in the population of the lowest magnon band, the $\nu$ band, and consequently these thermally excited magnons renormalize the the $\nu$ band more strongly than the $\mu$ band. This can be readily seen in Figs. \ref{fig:biaxial-meanfield} and \ref{fig:biaxial-meanfield-hexagonal}.

\section{Thermal Squeezed Magnons} \label{polarization}
\begin{figure}[h]
    \includegraphics[width=80mm]{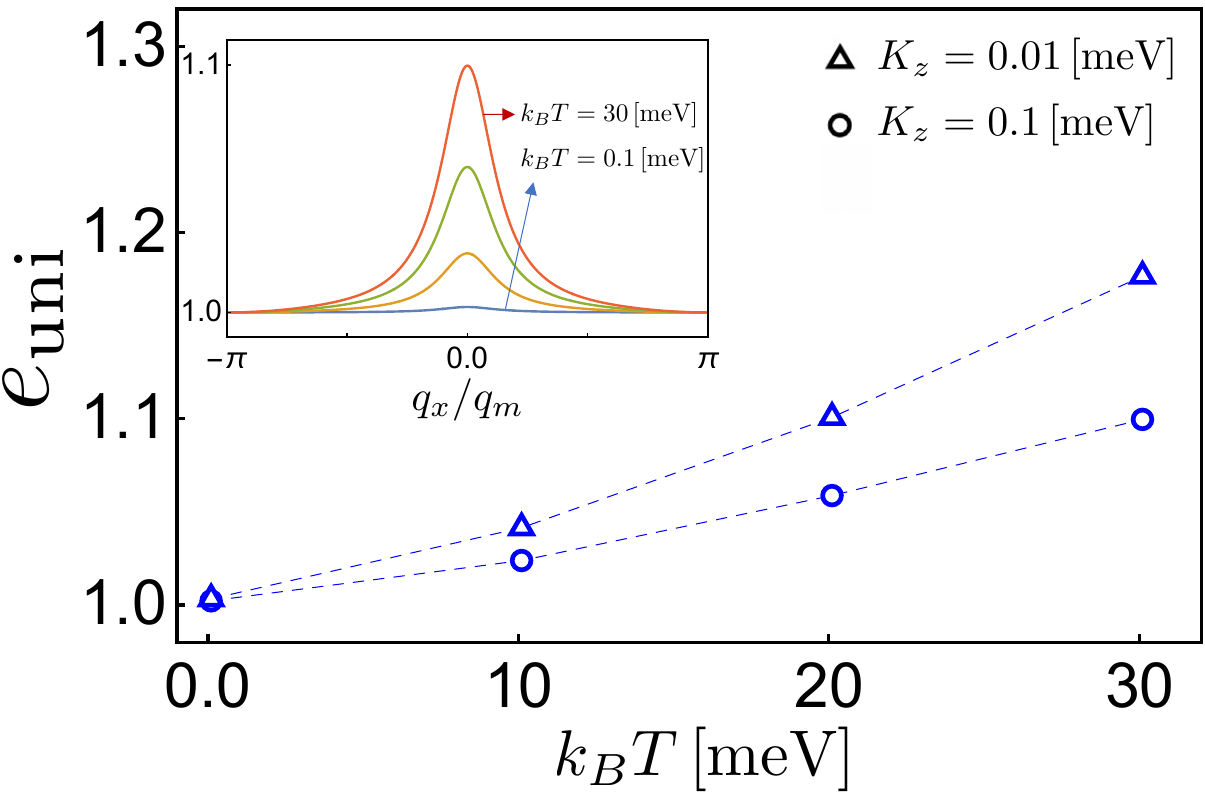}
    \caption{Temperature-dependent ellipticity of AF resonant modes of a uniaxial AF square lattice for different easy-axis anisotropy ($K_z$). The two AF modes are degenerate with the same ellipticity $e_{\rm{uni}}$. The inset shows ellepticity vs. dimensionless wavenumber for different temperatures, $k_BT=\{0.1, 10, 20, 30\}$ [meV].}\label{fig: ellipticity-uni}
\end{figure}

\begin{figure}[h]
    \includegraphics[width=\columnwidth]{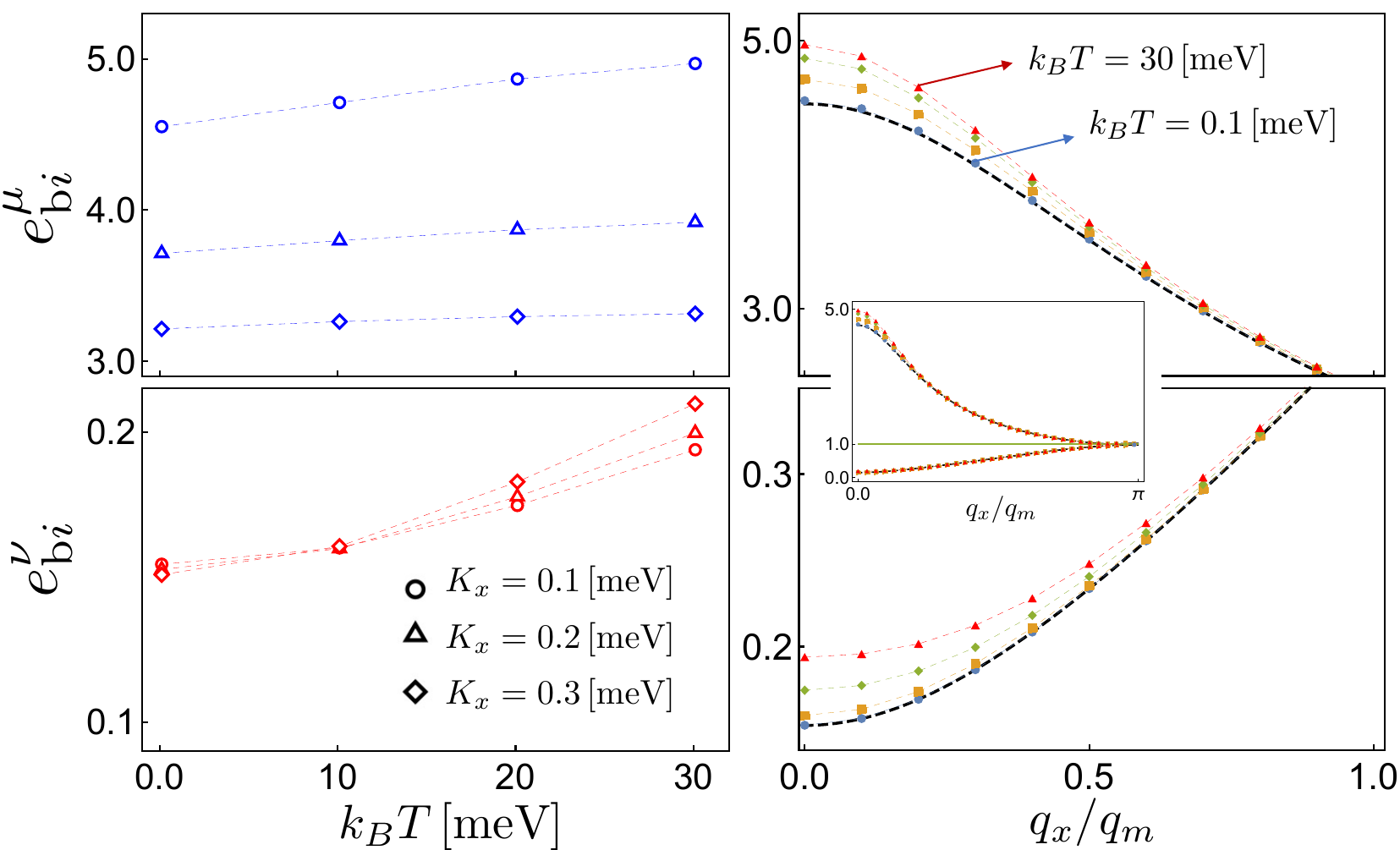}
    \caption{Ellipticity of two AF resonance modes as a function of   temperature (left panel) and dimensionless wavenumber (right panel) in a biaxial AF square lattice. The temperature-dependent ellipticity of each AF resonance mode, $e^{\mu}_{\rm{bi}}$ and $e^{\nu}_{\rm{bi}}$, are represented at left panels by blue and red curves, respectively. The result is shown for various hard-axis  anisotropies, $K_x$, with a fixed easy axis anisotropy $K_z=0.1$. Right panels show the  wavenumber dependence of $e^{\mu}_{\rm{bi}}$ and $e^{\nu}_{\rm{bi}}$ for different temperatures, $k_BT=\{0, 0.1, 10, 20, 30\}$ [meV] around the center of the first BZ. Black dashed line represents the ellipticity of noninteracting magnons. The inset in the right panel shows the behaviour of ellipticity in the whole BZ, where two magnon modes become circularly polarized close to the first BZ boundaries.}\label{fig: ellipticity-biaxial}
\end{figure}

In this section, we study the effect of nonlinear thermal magnons on the polarization of magnon eigenmodes based on the calculation in Appendix \ref{app: elipticity}. In general, magnon polarization is an essential ingredient in magnonic-based information technology \cite{PhysRevB.97.020402,PhysRevB.104.054419,Barman_2021}. Therefore, it is important to understand the effect of interactions on the polarization of magnon eigenmodes.  As we have mentioned earlier, two-sublattice AF systems have two magnon eigenmodes. At long-wavelength limit, i.e., close to the magnetic Brillouin zone center, the polarization of magnons is determined by magnetic anisotropy, while at short-wavelength, i.e.,  close to the magnetic Brillouin zone boundaries, the polarization of the modes are governed by the Heisenberg exchange interactions.
Therefore, in our spin model with isotropic Heisenberg exchange interaction in the presence of either uniaxial or biaxial magnetic anisotropies, high energy magnon eigenmodes close to the BZ magnetic boundaries are circularly polarized with an ellipticity $e=\langle S^x_i\rangle_{\rm{max}}/\langle S^y_i\rangle_{\rm{max}}=1$, see the inset of Fig. \ref{fig: ellipticity-uni} and right panel of Fig. \ref{fig: ellipticity-biaxial}. On the other hand, in the previous section, we have shown that nonlinear thermal interactions mostly modify the low-energy magnon excitations. Therefore, we expect interactions mainly affect on the polarization of the anisotropy-dominated low-energy magnons.

It has been shown, both theoretically and experimentally, that in quantum AF systems with negligible magnetic anisotropy and at low temperature, there are magnon self-squeezing states because of AF quantum fluctuations and interaction between two AF sublattices \cite{Peng2001-EPL,PhysRevLett.93.107203,Zhao2006PRB,Bossini,PhysRevB.105.054406}. Magnon squeezed states and quantum spin entanglement are related, and thus they are interesting for applications in quantum computing. In the presence of anisotropy and finite temperature, quantum magnon squeezing is reduced, but here we show that thermal squeezing becomes dominant at finite temperature \cite{cheng2008EPJ}. This thermal squeezing in uniaxial AF systems, arising from an effective attractive magnon interaction and breaking of the local U(1) symmetry, see Eq. (\ref{effectivH4uni}) and discussions below it.

The resulting temperature-dependent magnonic modes exhibit anisotropic spatial oscillations. This property is evidenced by the ellipticity, which is determined\cite{RezendeJAP2019} from the Bogolioubov transformation, (Eqs. \ref{BogoTuni} and \ref{bogoTbi}), and diagonalization matrix of the Hamiltonian $H={H}^{(2)}_{\text{m}}+{H}^{(4)}_{\text{MF}}$, that in turn relates the spin components and diagonal basis, $\hat{\mu}$ and $\hat{\nu}$, of operators, see \ref{app: elipticity} for more details. The result is obtained for the uniaxial ($e_{\rm{uni}}$) and biaxial ($e_{\rm{bi}}$) AF square-lattice systems, and presented at Figs. \ref{fig: ellipticity-uni} and \ref{fig: ellipticity-biaxial} as a function of temperature for different magnetic anisotropy strengths. Similar results is also obtained in hexagonal lattices (data not shown).
In Fig. \ref{fig: ellipticity-uni}, we plot the ellipticity $e_{\rm{uni}}$ for the degenerate magnon modes in uniaxial AF systems. The result is identical for each magnon mode, which is shown for various easy-axis anisotropy values. Interestingly, the eigenmode oscillations become squeezed along $y$-axis when temperature increases. This effect is a direct consequence of U(1) symmetry breaking arising from nonlinear thermal interactions, see Eq. (\ref{effectivH4uni}), and thus is termed {\it thermal squeezed magnons}\cite{cheng2008EPJ}, unlike its quantum counterpart defined at $T=0$ \cite{Peng2001-EPL,PhysRevLett.93.107203,Zhao2006PRB,Bossini}. In the biaxial AF system the ellipticity, $e_{\rm{bi}}$, of $\mu$- and $\nu$-magnonic eigenmode, are plotted at Fig. \ref{fig: ellipticity-biaxial} as a function of temperature and momentum. In this case, the eigenmodes are nondegenerate and they are elliptical even at zero temperature along different directions. The ellipticities, $e^{\mu}_{\rm{bi}}$ and $e^{\nu}_{\rm{bi}}$, at ${\bs q}={\bs 0}$ are represented (left panels) by blue and red curves, respectively. The magnonic $\mu$($\nu$)-mode becomes more (less) elliptical, being $e^{\mu}_{\rm{bi}}>1$ and $e^{\nu}_{\rm{bi}}<1$, when temperature increases but in an anisotropic way. Similarly, the dependence with momentum is displayed at right panels. As temperature is increased, both $e^{\mu}_{\rm{bi}}$ and $e^{\nu}_{\rm{bi}}$, slightly deviate from the ellipticity for noninteracting magnons (i.e., at $T=0$ and depicted by a black dashed line). Different rate in the change of ellipticity of the two modes vs. temperature is related to the fact that the $\nu$ mode has lower energy and this band becomes more populated than the $\mu$ mode; therefore the nonlinear effects are stronger in this lower band, see Fig. \ref{fig:biaxial-meanfield}.

\section{Summary and Conclusion}\label{sec:conclusion}
We have developed a self-consistent Hartree-Fock mean-field formalism to compute nonlinear magnon interactions in two-sublattice biaxial AF systems at finite temperature. We have shown that thermal interactions, on the one hand, reduce magnon bandgaps and thus AF resonance modes. Magnon bandgap plays an important role in the stability of long-range magnetic order in 2D systems. On the other hand, we have investigated the effect of nonlinear thermal magnon interactions on the polarization of low-energy magnons. We have shown that in the uniaxial AF case, magnon modes become squeezed because of the attractive nature of nonlinear magnon interactions and the breaking of local U(1) symmetry.
In the present paper, we have used the truncated HP bosonization transformation scheme to compute nonlinear magnons. The Dyson-Maleev transformation \cite{PhysRev.102.1217,Maleev} also leads to the same effective renormalized Hamiltonian within the Hartree-Fock approximation \cite{Liu_1992, PhysRevB.104.064435}.
We have shown that long-wavelength magnons interact more strongly than high-energy magnons. Even at zero temperature, there is a finite nonlinear quantum correction to the bare magnon dispersion. The power of this method is that it not only simplifies the interaction terms in the magnon Hamiltonian but also does not impose a new constraint on the results. The HP bosonization method is applied to find magnon interactions, and the Hartree-Fock method, applied to find mean-field thermal magnon interactions, is more accurate at low magnon numbers and large spin limits. Within this formalism, it is easy to investigate the effect of nonlinear magnons in various magnonic phenomena such as ultrafast magnon excitation and transport, two-magnon dynamics, and magnon condensation.

\acknowledgements
M. Sh. acknowledges Dr. Morteza Mohseni for fruitful discussions. M. Sh. and J. F. acknowledge support from Carl Tryggers Stiftelse, and J. F. acknowledges support from Vetenskapsr\aa det. 
This project has been supported by the Norwegian Financial Mechanism Project No. 2019/34/H/ST3/00515, ``2Dtronics''; and partially by the Research Council of Norway through its Centres of Excellence funding scheme, Project No. 262633, ``QuSpin''.

\bibliographystyle{apsrev4-1}
\bibliography{ref.bib}

\clearpage
\onecolumngrid
\appendix 

\section{Mean-field Coefficients}\label{appB}
In this Appendix, we present the treatment of the magnon-magnon Hamiltonian $H^{(4)}_b$ within the mean-field approximation. 
We define the following mean-field parameters\cite{balucani1980magnetic},
\begin{subequations}\label{meanfieldpara}
\begin{align}
&
m
= \frac{2}{N S} \sum_{\bs q} \langle a^\dag_{\bs q} a_{\bs q} \rangle = \frac{2}{N S}\sum_{\bs q}  \langle b^\dag_{\bs q} b_{\bs q} \rangle,
\\&
m^\prime = \frac{2}{N S} \sum_{\bs q} \gamma_{-{\bs q}} \langle a_{\bs q} b^\dag_{\bs q} \rangle,
\\&
p = \frac{2}{N S} \sum_{\bs q} \langle a_{\bs q} a_{-{\bs q}} \rangle = \frac{2}{N S}\sum_{\bs q} \langle b_{\bs q} b_{-{\bs q}} \rangle,
\\&
p^\prime = \frac{2}{N S} \sum_{\bs q} \gamma_{-{\bs q}} \langle a_{\bs q} b_{-{\bs q}} \rangle,
\end{align}
\end{subequations}
where $\langle...\rangle$ is for the thermal average. The mean-field parameter $m$ denotes the number of bosonic excitations on each sublattice $\cal{A}$ and $\cal{B}$, while $m'$, $p$ and $p'$ denote the interaction between them. In the following, we evaluate these mean-field parameters for both  uniaxial and biaxial cases, separately.

\subsection{Uniaxial easy-axis AF system}\label{appBuni}
In the uniaxial AF case, where the magnon basis reads as Eq. (\ref{BogoTuni}), the mean-field parameters become,
\begin{subequations}\label{meanfieldparauni}
\begin{align}
m&=\frac{2}{N S} \sum_{\bs q} \big(|u^u_{\bs q}|^2+|v^u_{\bs q}|^2\big) \, n_{\bs q} + |v^u_{\bs q}|^2,
\\
m^\prime &= p=0,
\\
p^\prime& = -\frac{2}{N S} \sum_{\bs q} \gamma_{-{\bs q}} u^u_{\bs q} v^u_{\bs q} \, (2 n_{\bs q} + 1 ).
\end{align}
\end{subequations}

The mean-field coefficients, introduced in the Hamiltonian (\ref{effectivH4uni}) obey,
\begin{subequations}
\begin{align}
{\cal C}_{\bs q}& = \frac{1}{2} [A_1 (|u^u_{\bs q}|^2 + |v^u_{\bs q}|^2)- (A_{2, \bs q} \,  \bar{u}^u_{\bs q} \bar{v}^u_{\bs q} + \bar{A}_{2, \bs q} \, u^u_{\bs q} v^u_{\bs q})],
\\
{\cal C}^\prime_{\bs q}& = \frac{1}{2} [A_1 (|u^u_{\bs q}|^2 + |v^u_{\bs q}|^2)- (A_{2, -\bs q} \, \bar{u}^u_{\bs q} \bar{v}^u_{\bs q} + \bar{A}_{2, -\bs q} \, u^u_{\bs q} v^u_{\bs q})],
\\
{\cal D}^\prime_{\bs q}& =
-2 A_1 \bar{u}^u_{\bs q} v^u_{\bs q} + (\bar{A}_{2, \bs q} (v^u_{\bs q})^{2} + A_{2, \bs q} (\bar{u}^u_{\bs q})^2).
\end{align}
\end{subequations}
where $A_1 = - \big(z J + 4 \mathcal{K}_z\big) S m - z J S (p^\prime+\bar{p}^\prime)/2$ and $A_2 = - z J S (\bar{p}^\prime+m) \, \gamma_{-\bs q}$, and $n_{\bs q}=(n_{\alpha,\bs q}+n_{\beta,\bs q})/2$. In the uniaxial case and in the absence of magnetic fields and DMIs $n_{\alpha,\bs q}=n_{\beta,\bs q}$.

\subsection{Biaxial easy-plane AF system}\label{appBbi}
In this case ${\cal K}_x\neq 0$, where the magnon basis reads as Eq. (\ref{bogoTbi}), the mean-field parameters become, 
\begin{subequations}
\begin{align}
m&\label{eq:meanfield-m}=\frac{2}{N S} \sum_{\bs q}\left[
\big(|u_{\bs q}|^2 + |x_{\bs q}|^2 \big) \, n_{\alpha, {\bs q}} + \big( |v_{\bs q}|^2 + |w_{\bs q}|^2 \big) \, n_{\beta, {\bs q}}+|x_{\bs q}|^2 + |w_{\bs q}|^2\right],
\\
p&\label{eq:meanfield-p}=\frac{2}{N S} \sum_{\bs q}\left[
u_{\bs q} x_{\bs q} \, \big( 2 n_{\alpha, {\bs q}} + 1 \big) + v_{\bs q} w_{\bs q} \, \big( 2 n_{\beta, {\bs q}} + 1 \big)\right],
\\
m^\prime&\label{eq:meanfield-m'}=\frac{2}{N S} \sum_{\bs q} \gamma_{-{\bs q}} \left[-\big( |u_{\bs q}|^2 + |x_{\bs q}|^2 \big) \, n_{\alpha, {\bs q}} + \big( |v_{\bs q}|^2 + |w_{\bs q}|^2 \big) \, n_{\beta, {\bs q}}
- |u_{\bs q}|^2 + |v_{\bs q}|^2\right],
\\
p^\prime&\label{eq:meanfield-p'}=\frac{2}{N S}\sum_{\bs q} \gamma_{-{\bs q}} \left[
- u_{\bs q} x_{\bs q} \, \big( 2 n_{\alpha, {\bs q}} + 1 \big) + v_{\bs q} w_{\bs q} \, \big( 2 n_{\beta, {\bs q}} + 1 \big)\right],
\end{align}
\end{subequations}
which depend on the Bose-Einstein distribution function of magnons at each mode, $n_{\alpha,{\bs q}}$ and $n_{\beta,{\bs q}}$, and the Bogoliubov coefficients. In particular, note that the mean-field parameters are finite even at zero temperature. In this regime, the magnon distributions $ n_{\chi,{\bs q}}\rightarrow 0$, and thus $m = 2\sum_{\bs q}(|x_{\bs q}|^2 + |w_{\bs q}|^2)/NS$, $p = 2 \sum_{\bs q}
(u_{\bs q} x_{\bs q} + v_{\bs q} w_{\bs q})/NS$, $m^\prime = 2\sum_{\bs q} \gamma_{-{\bs q}}
(- |u_{\bs q}|^2 + |v_{\bs q}|^2)/NS$ and $p^\prime = 2 \sum_{\bs q} \gamma_{-{\bs q}}
(- u_{\bs q} x_{\bs q} + v_{\bs q} w_{\bs q})/NS$.
This results emphasize that magnon-magnon scattering introduces finite corrections in the magnon dispersion even at zero temperature.
\\
The mean-field coefficients, introduced in the Hamiltonian (\ref{eq:matrixH}) obey,
\begin{subequations}\label{Cs}
\begin{align}
\cal{C}_{\alpha\alpha,{\bs q}}=&\label{caa}
\frac{1}{2}\big[2 A_{a}( |u_{\bs q}|^2 +  |x_{\bs q}|^2) 
-
(A^\prime_{ab, {\bs q}} + \bar{A}^\prime_{ab, {\bs q}} ) \, |u_{\bs q}|^2
-
(A^\prime_{ab, {-\bs q}} + \bar{A}^\prime_{ab, {-\bs q}} ) \, |x_{\bs q}|^2
\\& \nonumber
+ ( 2 A^\prime_{a} + 2 A^\prime_{b} - A_{ab, \bs q} - A_{ab, -\bs q} ) \, u_{\bs q} x_{\bs q}
+ ( 2 \bar{A}^\prime_{a} + 2 \bar{A}^\prime_{b} - \bar{A}_{ab, \bs q} - \bar{A}_{ab, -\bs q} ) \, \bar{u}_{\bs q} \bar{x}_{\bs q}\big],
\\ \nonumber
\\
{\cal C}_{\beta\beta,{\bs q}} = &\label{cbb}
\frac{1}{2}\big[2 A_{a} ( |v_{\bs q}|^2 + |w_{\bs q}|^2)
+
(A^\prime_{ab, {\bs q}} + \bar{A}^\prime_{ab, {\bs q}} ) \, |v_{\bs q}|^2
+
(A^\prime_{ab, {-\bs q}} + \bar{A}^\prime_{ab, {-\bs q}} ) \, |w_{\bs q}|^2
\\& \nonumber
+ ( 2 A^\prime_{a} + 2 A^\prime_{b} + A_{ab, \bs q} + A_{ab, -\bs q} ) \, v_{\bs q} w_{\bs q}
+ ( 2 \bar{A}^\prime_{a} + 2 \bar{A}^\prime_{b} + \bar{A}_{ab, \bs q} + \bar{A}_{ab, -\bs q} ) \, \bar{v}_{\bs q} \bar{w}_{\bs q}\big],
\\ \nonumber
\\
\cal{C}^\prime_{\alpha\alpha,{\bs q}}=&\label{cpaa}
( 2 A_a - A^\prime_{ab, {\bs q}} - {\bar{A}^\prime}_{ab, {\bs q}}) u_{\bs q} \bar{x}_{\bs q}
+
( A^\prime_{a} + A^\prime_{b} - A_{ab, {\bs q}} ) u^2_{\bs q}
+
( {\bar{A}^\prime_{a}} + {\bar{A}^\prime_{b}} - \bar{A}_{ab, {\bs q}} ) {\bar{x}}^2_{\bs q},
\\ \nonumber
\\
\cal{C}^\prime_{\beta\beta,{\bs q}}=&\label{cpbb}
( 2 A_a + A^\prime_{ab, {\bs q}} + {\bar{A}^\prime}_{ab, {\bs q}}) v_{\bs q} \bar{w}_{\bs q}
+
( A^\prime_{a} + A^\prime_{b} + A_{ab, {\bs q}} ) v^2_{\bs q}
+
( {\bar{A}^\prime_{a}} + {\bar{A}^\prime_{b}} + \bar{A}_{ab, {\bs q}} ) {\bar{w}}^2_{\bs q},
\\ \nonumber
\\
\cal{D}_{\alpha\beta,{\bs q}}=&\label{Dab}
(A_{ab, -{\bs q}} - A_{ab, {\bs q}} )x_{\bs q} v_{\bs q}
-
(\bar{A}_{ab, -{\bs q}} - \bar{A}_{ab, {\bs q}} ) \bar{u}_{\bs q} \bar{w}_{\bs q}
+
\big( A^\prime_{ab, -{\bs q}} - {\bar{A}^\prime}_{ab,-{\bs q}} \big)x_{\bs q} \bar{w}_{\bs q}
-
\big( A^\prime_{ab, {\bs q}} - {\bar{A}^\prime}_{ab, {\bs q}} \big)v_{\bs q}\bar{u}_{\bs q}
\\& \nonumber
+
2 ( A^\prime_a - A^\prime_b ) v_{\bs q} x_{\bs q} 
+
2 ( \bar{A}^\prime_a - \bar{A}^\prime_b ) \bar{u}_{\bs q} \bar{w}_{\bs q},
\\ \nonumber
\\
\cal{D}^\prime_{\alpha\beta,{\bs q}}=&\label{Dpab}
(\bar{A}_{ab, -{\bs q}} - \bar{A}_{ab, {\bs q}})\bar{x}_{\bs q}\bar{w}_{\bs q}
+
(A_{ab,{\bs q}}-A_{ab, -{\bs q}})u_{\bs q}v_{\bs q}
+
\big( { \bar{A}'_{ab, -{\bs q}}} - A^\prime_{ab, -{\bs q}} \big)\bar{x}_{\bs q}v_{\bs q}
+
\big( A^\prime_{ab, {\bs q}} - {\bar{A}^\prime}_{ab, {\bs q}} \big) u_{\bs q} \bar{w}_{\bs q}
\\& \nonumber
+
2 ( A^\prime_a - A^\prime_b ) u_{\bs q} v_{\bs q} 
+
2 ( \bar{A}^\prime_a - \bar{A}^\prime_b ) \bar{x}_{\bs q} \bar{w}_{\bs q},
\end{align}
\end{subequations}
where the following definitions have been introduced
\begin{subequations}
\begin{align}
A_{a}&\label{eq:Aa} =  
- S \left[ z J
\left( m+\frac{p^\prime+\bar{p}^\prime}{2} \right) + 4 m \left( \mathcal{K}_z+\frac{\mathcal{K}_x}{2} \right) - \frac{3}{4} \left( p+\bar{p} \right) \mathcal{K}_x
\right]
\\
A^\prime_{a}&\label{eq:A'a}= -  S\left[ z J \bar{m}^\prime + \bar{p} \left(\mathcal{K}_z + \frac{\mathcal{K}_x}{2}\right) - \frac{3}{4} m \mathcal{K}_x \right],
\\
A^\prime_{b}&\label{eq:A'b} = -  S\left[z J  {m^\prime} +  \bar{p} \left(\mathcal{K}_z + \frac{\mathcal{K}_x}{2}\right) - \frac{3}{4} m \mathcal{K}_x \right],
\\
A_{ab, {\bs q}}&\label{eq:Aab} = -  z J S \, \gamma_{-{\bs q}} \, \left( \bar{p}^\prime + m \right)
\\
A^\prime_{ab, {\bs q}}&\label{eq:A'ab} = -  z J S \, \gamma_{-{\bs q}} \, \left(\bar{m}^\prime + \frac{p + \bar{p}}{4}\right).
\end{align}
\end{subequations}
In the cases evaluated in the main text, i.e., in the absence of magnetic field and DMIs, the Bogoliubov coefficients $u_{\bs q}, v_{\bs q}, w_{\bs q}$ and $x_{\bs q}$ can be defined as real functions, thus the previous relations for the mean-field coefficients are simplified in the following form,
\begin{subequations}\label{simplecoeffHbi}
\begin{align}
{\cal C}_{\alpha\alpha,{\bs q}} & = \Delta^{-}_{\bs q} |u_{\bs q}|^2 + \Delta^{-}_{-\bs q}|x_{\bs q}|^2 + 2 \, \text{Re} \left[ u_{\bs q} x_{\bs q} \, \big(\xi - \text{Re} \, \Gamma_q  \big) \right],
\\
\\ \nonumber
{\cal C}_{\beta\beta,{\bs q}} &= \Delta^{+}_{\bs q} |v_{\bs q}|^2 + \Delta^{+}_{\bs q} |w_{\bs q}|^2 +  2 \, \text{Re} \left[ v_{\bs q} w_{\bs q} \, \big(\xi + \text{Re} \, \Gamma_q  \big) \right],
\\
\\ \nonumber
{\cal C}^\prime_{\alpha\alpha,{\bs q}}&=2\Delta^{-}_{\bs q}u_{\bs q}{x}_{\bs q}+\Lambda^{-}_{\bs q} u^2_{\bs q}+ \bar{\Lambda}^{-}_{\bs q}{x}^2_{\bs q},
\\
\\ \nonumber
{\cal C}^\prime_{\beta\beta,{\bs q}}&=2\Delta^{+}_{\bs q}v_{\bs q}{w}_{\bs q}+ \Lambda^{+}_{\bs q} v^2_{\bs q}+ \bar{\Lambda}^{+}_{\bs q} {w}^2_{\bs q},
\\
\\ \nonumber
{\cal D}_{\alpha\beta,{\bs q}}&=\Gamma_{\bs q}x_{\bs q}v_{\bs q}-\bar{\Gamma}_{\bs q}{u}_{\bs q}{w}_{\bs q}+\Theta_{-\bs q}x_{\bs q}{w}_{\bs q}-\Theta_{\bs q}v_{\bs q}{u}_{\bs q} + 2 \, \xi v_{\bs q} \, x_{\bs q} + 2 \, \bar{\xi} \, u_{\bs q} \, w_{\bs q},
\\
\\ \nonumber
{\cal D}^\prime_{\alpha\beta,{\bs q}}& = \bar{\Gamma}_{\bs q}{x}_{\bs q}{w}_{\bs q} - \Gamma_{\bs q} u_{\bs q} v_{\bs q} - {\Theta}_{-\bs q}{x}_{\bs q} v_{\bs q} + \Theta_{\bs q}{w}_{\bs q} u_{\bs q} + 2 \xi^\prime u_{\bs q} v_{\bs q} + 2 \, \bar{\xi}^\prime \, x_{\bs q} w_{\bs q}.
\end{align}
\end{subequations}
where we define,
\begin{subequations}\label{simpcoeff}
\begin{align}
\Delta^{\pm}_{\bs q} & = A_a \pm \text{Re}\left[A^\prime_{ab, {\bs q}}\right],
\\
\xi & = A_a^\prime + A_b^\prime,
\hspace{3cm}
\xi^\prime = A_a^\prime - A_b^\prime,
\\
\Gamma_{\bs q} & = (A_{ab,-{\bs q}}-A_{ab,{\bs q}}),
\hspace{2cm}
\Gamma_{\bs q}^\prime = (A_{ab,-{\bs q}}+A_{ab,{\bs q}}),
\\
\Lambda^{\pm}_{\bs q} & = \xi \pm A_{ab, {\bs q}},
\\
\Theta_{\bs q} & = (A'_{ab,{\bs q}}-\bar{A}'_{ab,{\bs q}}).
\end{align}
\end{subequations}

\subsection{Eigenenergies of a biaxial easy-plane AF system}\label{app:eigenenergiesbiaxial}
\begin{figure}[htp]
    \includegraphics[width=120mm]{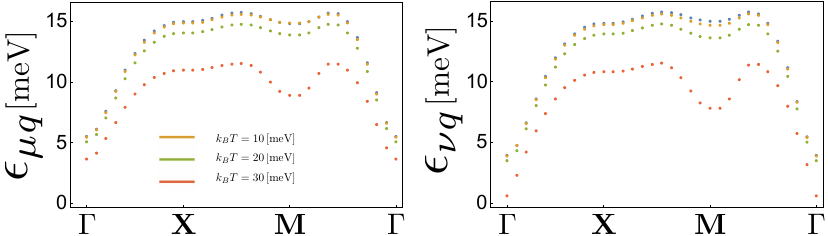}
    \caption{Dispersion relation along the high-symmetry lines for temperature-dependent magnonic eigenmodes of a 3D cubic biaxial AF insulator. The eigenenergies for each mode, $\epsilon_{\mu}$ (left panel) and $\epsilon_{\nu}$ (right panel), are presented for various temperatures, using Eq. (\ref{eq:magnonenergies-biaxial}).}\label{fig: 3DcubicbiaxialAF}
\end{figure}

The eigenenergy of a biaxial AF Hamiltonian in the presence of nonlinear interaction corrections, modelled by an effective mean-field Hamiltonian $H=H^{(2)}_{\text{m}}+H^{(4)}_{\text{MF}}$, where $H=H^{(2)}_{\text{m}}$ is given by Eq. (\ref{noninteracting}) and $H^{(4)}_{\text{MF}}$ is given by Eq. (\ref{eq:matrixH}); are given by,
\begin{align}\label{eq:magnonenergies-biaxial}
2\tilde{\epsilon}^2_{\mu,\nu}\nonumber={\cal X}&\pm\left[{{\cal X}^2-\left({\cal D}^{\prime2}_{\alpha\beta} \left(0.25\bar{{\cal D}}^{\prime2}_{\alpha\beta}-\bar{\cal C}^{\prime}_{\alpha\alpha}\bar{\cal C}^{\prime}_{\beta\beta}\right)+ 4\left(\epsilon_{\alpha}+0.5{\cal C}_{\alpha\alpha}\right)^2 \left(\left(\epsilon_{\beta}+0.5{\cal C}_{\beta\beta}\right)^2 - {\cal C}^{\prime}_{\beta\beta} \bar{\cal C}^{\prime}_{\beta\beta}\right)\right)}\right.\\
&\nonumber\left.+{\cal C}^{\prime}_{\alpha\alpha} \left(4\left(\epsilon_{\beta}+0.5{\cal C}_{\beta\beta}\right)^2 \bar{\cal C}^{\prime}_{\alpha\alpha} + \bar{{\cal D}}_{\alpha\beta}^{\prime2} {\cal C}^{\prime}_{\beta\beta} -4 \bar{\cal C}^{\prime}_{\alpha\alpha} |{\cal C}^{\prime}_{\beta\beta}|^2 - 2 \left(\epsilon_{\beta}+0.5{\cal C}_{\beta\beta}\right) \bar{{\cal D}}^{\prime}_{\alpha\beta}{\cal D}_{\alpha\beta} + \bar{\cal C}^{\prime}_{\beta\beta} {\cal D}_{\alpha\beta}^2\right)\right.\\
&\nonumber\left.  -2{\cal D}^{\prime}_{\alpha\beta}\bar{\cal D}_{\alpha\beta}\left(\left(\epsilon_{\beta}+0.5{\cal C}_{\beta\beta}\right) \bar{\cal C}^{\prime}_{\alpha\alpha} - 0.25\bar{{\cal D}}^{\prime}_{\alpha\beta}{\cal D}_{\alpha\beta}\right)  - \left(-\bar{\cal C}^{\prime}_{\alpha\alpha} {\cal C}^{\prime}_{\beta\beta} + 0.25{\cal D}^2_{\alpha\beta}\right) \bar{\cal D}^2_{\alpha\beta}\right.\\
&\left.-2 \left(\epsilon_{\alpha}+0.5{\cal C}_{\alpha\alpha}\right) \left({\cal D}^{\prime}_{\alpha\beta} \bar{\cal C}^{\prime}_{\beta\beta} {\cal D}_{\alpha\beta} + \bar{{\cal D}}^{\prime}_{\alpha\beta} {\cal C}^{\prime}_{\beta\beta} \bar{\cal D}_{\alpha\beta} - \left(\epsilon_{\beta}+0.5{\cal C}_{\beta\beta}\right) \left(|{\cal D}^{\prime}_{\alpha\beta}|^2+ |{\cal D}_{\alpha\beta}|^2\right)\right)
\right]^{\frac{1}{2}},
\end{align}
with ${\cal X}=\left(\epsilon_{\alpha}+0.5{\cal C}_{\alpha\alpha}\right)^2+\left(\epsilon_{\beta}+0.5{\cal C}_{\beta\beta}\right)^2+0.5\left(\left|{\cal D}_{\alpha\beta}\right|^2-\left|{\cal D}'_{\alpha\beta}\right|^2\right)-\left|{\cal C}'_{\alpha\alpha}\right|^2-\left|{\cal C}'_{\beta\beta}\right|^2$.
In square lattices, as we have already discussed in the main text, the interband scattering processes are absent ${\cal D}_{\alpha\beta}={\cal D}'_{\alpha\beta}=0$, and thus Eq. (\ref{eq:magnonenergies-biaxial}) reduces to Eq. (\ref{E-sq}) in the main text. The result of Eq. (\ref{eq:magnonenergies-biaxial}), is computed for cubic lattices and presented at Fig. \ref{fig: 3DcubicbiaxialAF} for different temperatures.

\subsection{Ellipticity}\label{app: elipticity}
The ellipticity is defined as $e=\langle S^x_i\rangle_{\rm{max}}/\langle S^y_i\rangle_{\rm{max}}$.
In the uniaxial easy-axis AF case, the elipticity of each AF eigenmode is given by,
\begin{subequations}
\begin{align}
e^{\mu}_{\text{uni}}&=\frac{\sqrt{A+\varepsilon_{\bs q}}+\sqrt{A-\varepsilon_{\bs q}}}{\sqrt{A+\varepsilon_{\bs q}}-\sqrt{A-\varepsilon_{\bs q}}},\\
e^{\nu}_{\text{uni}}&=\frac{\sqrt{A'+\varepsilon_{\bs q}}+\sqrt{A'-\varepsilon_{\bs q}}}{\sqrt{A'+\varepsilon_{\bs q}}-\sqrt{A'-\varepsilon_{\bs q}}},
\end{align}
\end{subequations}
where $A=\epsilon^{\text{u}}_{\bs q}+0.5{\cal C}_{\bs q}$ and $A'=\epsilon^{\text{u}}_{\bs q}+0.5{\cal C}'_{\bs q}$. Since ${\cal C}_{\bs q}={\cal C}'_{\bs q}$, two modes have the same ellipticity. 

In the biaxial easy-plane AF system, we find the following relations for ellipticity of each eigenmode,
\begin{subequations}
\begin{align}
e^{\mu}_{\text{bi}}&=\frac{P_{11}+P_{13}}{P_{11}-P_{13}},\\
e^{\nu}_{\text{bi}}&=\frac{P_{14}+P_{12}}{P_{14}-P_{12}},
\end{align}
\end{subequations}
where we have defined the following parameters,
\begin{subequations}
\begin{align}
P_{11}&=u_{\bs q}Q_{11}+v_{\bs q}Q_{21}+x_{\bs q}Q_{13}+w_{\bs q}Q_{23},\\
P_{13}&=u_{\bs q}Q_{13}+v_{\bs q}Q_{23}+x_{\bs q}Q_{11}+w_{\bs q}Q_{21},\\
P_{14}&=u_{\bs q}Q_{12}+v_{\bs q}Q_{22}+x_{\bs q}Q_{14}+w_{\bs q}Q_{24},\\
P_{12}&=u_{\bs q}Q_{14}+v_{\bs q}Q_{24}+x_{\bs q}Q_{12}+w_{\bs q}Q_{22}.
\end{align}
\end{subequations}
The matrix $Q$ is the transformation  matrix that diagonalizes the total interacting Hamiltonian $H=H^{(2)}_{\text{m}}+H^{(4)}_{\text{MF}}$, where  $H^{(4)}_{\text{MF}}=H^{(4)}_{\text{intra}}+H^{(4)}_{\text{inter}}$ is the  mean-field interaction, see Eqs. \eqref{effectivH4intra}-\eqref{effectivH4inter} in the main text. In the diagonal basis, the Hamiltonian is written as
\begin{align}
H=\frac{1}{2}\sum_{\bs q}({\Upsilon})^{\dagger}\Omega\,({\Upsilon}),
\end{align}
where $({\Upsilon})={\bf Q}\,({\Psi})$ is the new basis in which the Hamiltonian is diagonal  $({\Upsilon})=\left(\mu_{\bs q},\nu_{\bs q},\mu^{\dagger}_{-\bs q},\nu^{\dagger}_{-\bs q}\right)^T$, while the old basis is $({\Psi})=\left(\alpha_{\bs q},\beta_{\bs q},\alpha^{\dagger}_{-\bs q},\beta^{\dagger}_{-\bs q}\right)^T$, and $\Omega$ is a $4\times4$ diagonal matrix, with elements equivalent to the eigenenergies of the problem. and
\begin{align}
{\bf Q}=\left(\begin{array}{cc}
{\bf Q}_1   & {\bf Q}_2 \\
\bar{\bf Q}_2  & \bar{\bf Q}_1 
\end{array}\right),\qquad
{\bf Q}_1=\left(\begin{array}{cc}
{Q}_{11}   & { Q}_{12} \\
{Q}_{21}  & {Q}_{22} 
\end{array}\right),\qquad
{\bf Q}_2=\left(\begin{array}{cc}
{Q}_{13}   & { Q}_{14} \\
{Q}_{23}  & {Q}_{24} 
\end{array}\right)
\end{align}
being the transformation matrix\cite{RezendeJAP2019}.
\end{document}